\documentclass{aa}  

\usepackage{graphicx}
\usepackage{color}
\usepackage{txfonts}
\begin{document} 
\def\ammo{\rm NH_3}
\def\dammo{\rm NH_2D}
\def\ddammo{\rm NHD_2}
\def\ndthree{\rm ND_3}
\def\diaz{\rm N_2H^+}
\def\ddiaz{\rm N_2D^+}
\def\hthree{\rm H_3^+}
\def\htwod{\rm H_2D^+}
\def\dtwoh{\rm D_2H^+}
\def\dthree{\rm D_3^+}
\def\ammohplus{\rm NH_4^+}
\def\dammohplus{\rm NH_3D^+}
\def\ddammohplus{\rm NH_2D_2^+}
\def\ndthreehplus{\rm NHD_3^+}
\def\ammoplus{\rm NH_3^+}
\def\dammoplus{\rm NH_2D^+}
\def\ddammoplus{\rm NHD_2^+}
\def\ndthreeplus{\rm ND_3^+}
\def\htwo{\rm H_2}
\def\el{\rm e^-}
\def\hplus{\rm H^+}

\def\Harpoons{\mathop{\rightleftharpoons}\limits}
\def\arrow{\mathop{\rightarrow}\limits}

\title{Deuteration of ammonia in the starless core Ophiuchus/H-MM1
  \thanks{Based on observations carried out with The Atacama
    Pathfinder Experiment (APEX), the Robert C. Byrd Green Bank
    Telescope (GBT), and the IRAM 30m Telescope. APEX is a
    collaboration between Max-Planck Institut f\"ur Radioastronomie
    (MPIfR), Onsala Space Observatory (OSO), and the European Southern
    Observatory (ESO).  GBT is managed by the National Radio
    Astronomy Observatory, which is a facility of the National Science
    Foundation operated under cooperative agreement by Associated
    Universities, Inc.  IRAM is supported by INSU/CNRS (France), MPG
    (Germany), and IGN (Spain).}}


\author{J.~Harju\inst{1,2} \and F.~Daniel\inst{3,4} \and
  O.~Sipil{\"a}\inst{1} \and P.~Caselli\inst{1} \and
  J.E.~Pineda\inst{1} \and R.K.~Friesen\inst{5} \and
  A.~Punanova\inst{1} \and R.~G{\"u}sten\inst{6} \and
  L.~Wiesenfeld\inst{3,4,1} \and P.C.~Myers\inst{7} \and
  A.~Faure\inst{3,4} \and P.~Hily-Blant\inst{3,4} \and
  C.~Rist\inst{3,4} \and E.~Rosolowsky\inst{8} \and
  S.~Schlemmer\inst{9} \and Y.L.~Shirley\inst{10} }
\institute{
  {Max-Planck-Institute for Extraterrestrial Physics (MPE),
    Giessenbachstr. 1, 85748 Garching, Germany, e-mail:
    \texttt{harju@mpe.mpg.de}}
  \and{Department of Physics, P.O. Box 64, 00014 University of
    Helsinki, Finland}
  \and{Universit{\'e} Grenoble Alpes, IPAG, F-38000 Grenoble, France}
  \and{CNRS, IPAG, F-38000 Grenoble, France}
  \and{Dunlap Institute for Astronomy and Astrophysics, University of
    Toronto, 50 St. George Street, Toronto M5S 3H4, Ontario, Canada}
  \and{Max-Planck-Institut f{\"u}r Radioastronomie, Auf dem H{\"u}gel
    69, 53121 Bonn, Germany}
  \and{Harvard-Smithsonian Center for Astrophysics, 60 Garden Street,
    Cambridge MA 02138, USA}
  \and{Department of Physics, 4-181 CCIS, University of Alberta,
    Edmonton, AB T6G 2E1, Canada}
  \and{I. Physikalisches Institut, Universit\"at zu K\"oln,
    Z\"ulpicher Strasse 77, 50937 K\"oln, Germany}
  \and{Steward Observatory, University of Arizona, 933 North Cherry
    Avenue, Tucson, AZ 85721, USA} 
}


   \date{Received ; accepted }

 
  \abstract
  {Ammonia and its deuterated isotopologues probe physical conditions in 
    dense molecular cloud cores. The time-dependence of
    deuterium fractionation and the relative abundances of different
    nuclear spin modifications are supposed to 
    provide means of determining the evolutionary
    stages of these objects.}
  {We aim to test the current understanding of spin-state chemistry of 
    deuterated species by determining the abundances and spin
    ratios of $\dammo$, $\ddammo$, and $\ndthree$ in a quiescent,
    dense cloud.}
  {Spectral lines of $\ammo$, $\dammo$, $\ddammo$, $\ndthree$, and
    $\ddiaz$ were observed towards a dense, starless core in Ophiuchus
    with the APEX, GBT, and IRAM 30-m telescopes. The
    observations were interpreted using a gas-grain chemistry model
    combined with radiative transfer calculations. The chemistry model
    distinguishes between the different nuclear spin states of light
    hydrogen molecules, ammonia, and their deuterated forms. Different
    desorption schemes can be considered.}
  {High deuterium fractionation ratios with $\dammo/\ammo \sim
    0.4$, $\ddammo/\dammo \sim 0.2$, and $\ndthree/\ddammo\sim 0.06$
    are found in the core. The observed {\sl ortho/para} ratios of
  $\dammo$ and $\ddammo$ are close to the corresponding nuclear spin
  statistical weights. The chemistry model can approximately reproduce
  the observed abundances, but predicts uniformly too low 
  {\sl ortho/para}-$\dammo$, and too large {\sl ortho/para}-$\ddammo$
    ratios. The longevity of $\diaz$ and $\ammo$ in dense gas, which
  is prerequisite to their strong deuteration, can be attributed to
  the chemical inertia of ${\rm N_2}$ on grain surfaces.}
  {The discrepancies between the chemistry model and the observations
    are likely to be caused by the fact that the model assumes
    complete scrambling in principal gas-phase deuteration reactions
    of ammonia, which means that all the nuclei are mixed in reactive
    collisions.  If, instead, these reactions occur through proton
    hop/hydrogen abstraction processes, statistical spin ratios are to
    be expected. The present results suggest that while the
      deuteration of ammonia changes with physical conditions and
      time, the nuclear spin ratios of ammonia isotopologues do not
      probe the evolutionary stage of a cloud. }


   \maketitle
%

\section{Introduction}

Ammonia belongs to the most useful probes of the dense cores of
molecular clouds owing to its energy spectrum and chemical properties
(\citealt{1983ARA&A..21..239H}; \citealt{1989ApJS...71...89B};
\citealt{2002ApJ...569..815T}; \citealt{2009ApJ...697.1457F}).  The
molecule can survive in the gas phase also in the cold, dense interior
parts of starless and prestellar cores, and in these regions reactions
with deuterated ions convert part of $\ammo$ to $\dammo$, $\ddammo$,
and $\ndthree$ (\citealt{2001ApJ...553..613R}). Ammonia and its
deuterated isotopologues are also formed on grain surfaces through
H/D-atom addition reactions to N atoms (\citealt{1989MNRAS.240P..25B};
\citealt{2015MNRAS.446..449F}).  Substantial deuteration in both
phases occurs after the disappearance of CO from the gas phase, and
needs some time to take effect, so the relative abundances of the
mentioned molecules can give an idea of the evolutionary stage of a
dense core (\citealt{2005A&A...438..585R};
\citealt{2006A&A...456..215F}; \citealt{2015A&A...576A..99R}).
Previously the $\ddiaz/\diaz$ abundance ratio has been used
extensively for this purpose (\citealt{2005ApJ...619..379C};
\citealt{2009A&A...494..623P}).

Besides enabling nuanced investigation into deuteration, spectral line
observations of $\ammo$, $\dammo$, $\ddammo$, and $\ndthree$ test our
understanding of spin-state chemistry, i.e., selection rules in
chemical reactions which determine the relative abundances of
different nuclear spin symmetry species or ``modifications''. Each
spectral line observed from these molecules belongs to one of the two
spin modifications of $\ammo$, $\dammo$, or $\ddammo$, or one of the
three spin modifications of $\ndthree$. The abundance ratios of these
modifications which must be treated as separate chemical species, can
deviate from their nuclear spin statistical ratios, and are predicted to
change with time (\citealt{2013ApJ...770L...2F};
\citealt{2015A&A...578A..55S}; \citealt{2015A&A...581A.122S}). 

The spin symmetry species of a molecule is defined based on how the
nuclear spin wave function transforms under symmetry operations like
the interchange or the permutation of identical nuclei. For molecules
containing only H nuclei there is a one to one correspondence between
the symmetry and the nuclear spin angular momentum, but this is no
more true for molecules with multiple D nuclei
\citep{2009JChPh.130p4302H}. \footnote{\citet{2016JChPh.145g4301S}
  have recently shown that the spin angular momentum and the
  permutation symmetry are, after all, inherently coupled, and that it
  is possible to construct an unequivocal and practical representation
  for each spin angular momentum - symmetry species using the Young
  diagrams. In the present study, chemical species are distinguished
  solely by their nuclear spin symmetries.}  The statistical weight of
a symmetry species is the number of possible nuclear spin functions
having that symmetry.  When there are no more than three spin
modifications, it is customary to call the species with the largest
nuclear spin statistical weight ``ortho'', and the one with the lowest
weight ``para''. If there is a third one, this is called ``meta''.
For example, the three elementary spin functions of the deuteron,
$|1,1\rangle$, $|1,0\rangle$, and $|1,-1\rangle$, can combine in
$\ndthree$ in 27 different ways, and these combinations can be
arranged to a set of 27 orthogonal functions, which form an
irreducible representations of the appropriate permutation group
$S_3$. Of these functions, 10 have symmetry $A_1$ (``meta''), 1 has
symmetry $A_2$ (``para''), and 16 have symmetry $E$ (``ortho'') in
$S_3$ (\citealt{BunkerJensen06}; \citealt{2015A&A...581A.122S};
\citealt{2016MNRAS.457.1535D}, see their Appendix A).

The abundances of all three deuterated forms of ammonia have been
previously estimated in the protostellar core ``B1-b'' of Barnard 1 in
Perseus (\citealt{2002ApJ...571L..55L}; \citealt{2005A&A...438..585R};
\citealt{2006ApJ...636..916L}), and in the starless core (``I16293E'')
of the dark cloud L1689N in Ophiuchus (\citealt{2005A&A...438..585R};
\citealt{2006A&A...454L..63G}). The fractionation ratios are similar
in these two objects. The recent analysis of these observations by 
\cite{2016MNRAS.457.1535D} gives $[\dammo]/[\ammo] \ga [\ddammo]/[\dammo]
\approx 0.2$, $[\ndthree]/[\ddammo] \approx 0.05-0.10$. In both cores,
the {\sl ortho/para} ratios of $\dammo$ and $\ddammo$ were found to be
close to their statistical values, 3:1 and 2:1, respectively. The
interpretation of chemical data from these two, bright sources of
deuterated ammonia is complicated by the possibility of enhanced
evaporation from grains because of shocks. The Barnard 1 core contains
two, albeit very young and low-mass protostars driving outflows (B1-bS
and B1-bN; \citealt{2007A&A...468.1009H};
\citealt{2014ApJ...789...50H}; \citealt{2015A&A...577L...2G}). The
core in L1689N may be compressed by an outlow from the adjacent
protostellar source IRAS 16293-2422 (\citealt{2002ApJ...569..322L};
\citealt{2004ApJ...608..341S}; \citealt{2006A&A...454L..63G}).

In the present work, we use the 100-m Robert C. Byrd Green Bank
telescope, the 12-m APEX, and the IRAM 30-m telescope to determine the
abundances of {\sl para}-$\ammo$, {\sl meta}-$\ndthree$, and both {\sl
  ortho} and {\sl para} modifications of $\dammo$ and $\ddammo$, in
the starless core H-MM1 in Ophiuchus (\citealt{2004ApJ...611L..45J};
\citealt{2011A&A...528C...2P}).  Through these observations we obtain
a homogenous data set pertaining to a quiescent region, presumably
characterised by a simple physical structure, where comparison with
chemistry models is more straightforward than in a star-forming
core. The observations are interpreted by modelling the chemical
evolution of a hydrostatic core resembling H-MM1, and simulating
observations towards this model core.  Besides the previously
published collisional coefficients of $\dammo$ with {\sl para}-$\htwo$
\citep{2014MNRAS.444.2544D}, we use newly calculated coefficients for
$\ddammo$ and $\ndthree$ \citep{2016MNRAS.457.1535D}, in conjunction
with the radiative transfer program of \cite{1997A&A...322..943J}.

The paper is organised as follows. In Sect.~2, we describe the target
core and give details of the observations. The direct observational
results are presented in Sect.~3. In Sect.~4, we construct a physical
model of the core and describe the modelling tools, the collisional
coefficients and the chemistry model, used in this work. In Sect.~5 we
make predictions for the $\ammo$, $\dammo$, $\ddammo$, and $\ndthree$
abundances, and the observable line emission from the model core. In
Sect.~6 we compare the outcome of the modelling with the observations,
and discuss the implications of this comparison. Finally, in Sect.~7
we draw our conclusions.

\section{Observations}

\subsection{The target: H-MM1} 

The dense core H-MM1 in Ophiuchus was discovered by
\cite{2004ApJ...611L..45J} using the Submillimeter Common-User
Bolometric Array (SCUBA) on the James Clerk Maxwell Telescope
(JCMT). The core was also covered by the JCMT Gould Belt Survey with
SCUBA-2 at 450 and 850 $\mu$m \citep{2015MNRAS.450.1094P}. H-MM1 lies
in relative isolation in the eastern part of Lynds 1688, far from
sites of active star formation. \cite{2011A&A...528C...2P} detected
extended {\sl para-}$\dtwoh$ emission towards this core. Based on the
analysis of the {\sl para}-$\dtwoh$ and {\sl ortho-}$\htwod$ lines
towards its centre, Parise et al. suggested that the average density
of the core is high compared to typical starless cores, a few
times $10^5$ cm$^{-3}$.  In accordance with the high abundances of
the deuterated ions, the deuterium fraction in $\diaz$ is also
extremely high: $\ddiaz/\diaz=0.43\pm0.11$ \citep{2016A&A...587A.118P}.

\begin{figure}
\includegraphics[width=9cm]{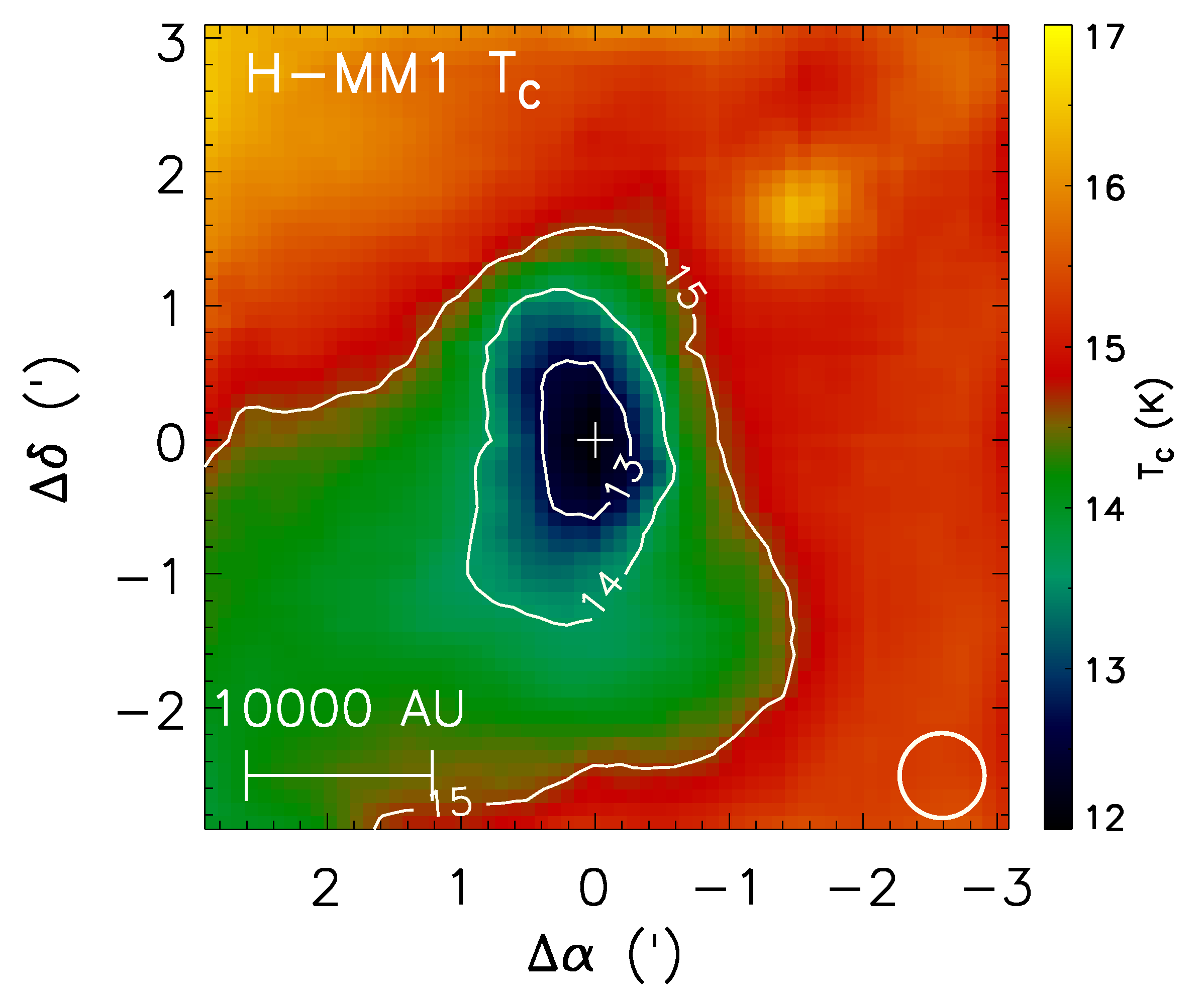}
\includegraphics[width=9cm]{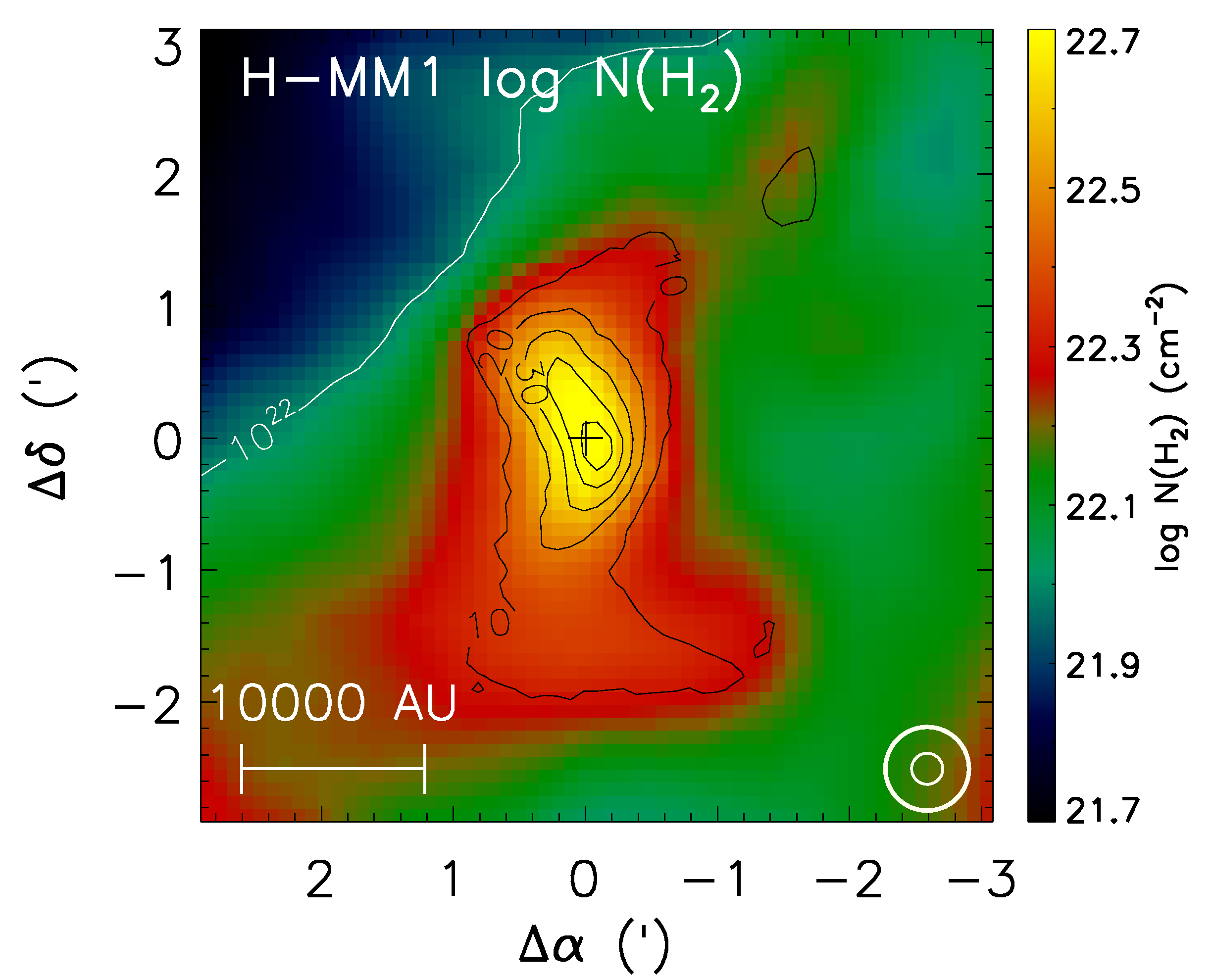}

\caption{Dust colour temperature ($T_{\rm C}$, top) and the $\htwo$
  column density ($N(\htwo)$, bottom) maps of H-MM1 as derived from
  Herschel/SPIRE maps at 250, 350, and 500 $\mu$m. The distribution of
  the 850 $\mu$m emission observed with SCUBA-2 is indicated with
  black contours on the $N(\htwo)$ map.  The contour levels are 10 to
  50 by 10 MJy\,sr$^{-1}$. The column density maximum is marked with a
  plus sign.  The present observations were pointed towards this
  position, with coordinates R.A. $16^{\rm h}27^{\rm m}59\fs0$,
  Dec. $-24\degr33\arcmin33\arcsec$ (J2000).  The larger circle in the bottom
  right represents the resolution of the $N(\htwo)$ and $T_{\rm C}$
  maps ($\sim 40\arcsec$). The $14^{\prime\prime}$ resolution of the
  SCUBA-2 $850\,\mu$m map is indicated with the smaller circle. The
  lowest $850\,\mu$m contour (10 MJy\,sr$^{-1}$) coincides roughly
  with the $N(\htwo)= 2\,10^{22}$ cm$^{-2}$ contour.}
\label{figure:hmm1_td_nh2_maps}
\end{figure}

In Fig.~\ref{figure:hmm1_td_nh2_maps} we show the dust colour
temperature ($T_{\rm C}$) and the $\htwo$ column density ($N(\htwo)$)
maps of the core derived from far-infrared images observed by {\sl
  Herschel} \citep{2010A&A...518L...1P}. Contours of the 850 $\mu$m emission 
map from the SCUBA-2 survey of \cite{2015MNRAS.450.1094P} are superposed 
on the $N(\htwo)$ map.  

The Herschel/SPIRE images were extracted from the pipeline-reduced
images of the Ophiuchus complex made in the course of the Herschel
Gould Belt Survey \citep{2010A&A...518L.102A}. The data are downloaded
from the Herschel Science Archive
(HSA)\footnote{\url{www.cosmos.esa.int/web/herschel/science-archive}}. We
calculated the $T_{\rm C}$ and $N(\htwo)$ distributions using only the
three SPIRE \citep{2010A&A...518L...3G} bands at $250\,\mu$m,
$350\,\mu$m, and $500\,\mu$m, for which the pipeline reduction
includes zero-level corrections based on comparison with the Planck
satellite data. A modified blackbody function with the dust emissivity
spectra index $\beta=2$ was fitted to each pixel, after smoothing the
$250\,\mu$m and $350\,\mu$m images to the resolution of the
$500\,\mu$m image ($\sim 40\arcsec$), and resampling all images to the
same grid. For the dust emissivity coefficient per unit mass of gas we
adopted the value from \cite{1983QJRAS..24..267H}, $\kappa_{250\mu
 \rm m}=0.1$ cm$^{2}$g$^{-1}$ ($1/C_{250}$ in Table I in their
paper). \cite{2013A&A...555A.140S} derived a similar value for
$\kappa_{250\mu \rm m}$ in the starless core CrA C. According to the
derived maps, the dust colour temperature minimum and the column
density maximum of the core is found at R.A. $16^{\rm h}27^{\rm
  m}59\fs0$, Dec. $-24\degr33\arcmin33\arcsec$ (J2000). The line observations
presented here were done towards this position. The obtained minimum
colour temperature is 12.4 K and the maximum column density is
$N(\htwo)=5.7\,10^{22}$ cm$^{-2}$. These values are averages over the
$40\arcsec$ beam. The fitted $T_{\rm C}$ overestimates the
mass-averaged dust temperature because of line-of-sight temperature
variations. This effect is more marked towards the centre of a starless core
than on the outskirts of the core (\citealt{2012A&A...547A..11N};
\citealt{2013A&A...555A.140S}).

Our position lies about $13\arcsec$ northeast from the centre position
used by \cite{2011A&A...528C...2P},  and about $7\arcsec$ east of 
the 450 and 850 $\mu$m peaks observed with SCUBA-2. In a later section we
will use the SCUBA-2 maps to derive a simple spherically symmetric
model of the core for the purpose of radiative transfer modelling.

\subsection{GBT observations}

\begin{figure}
\includegraphics[width=9cm]{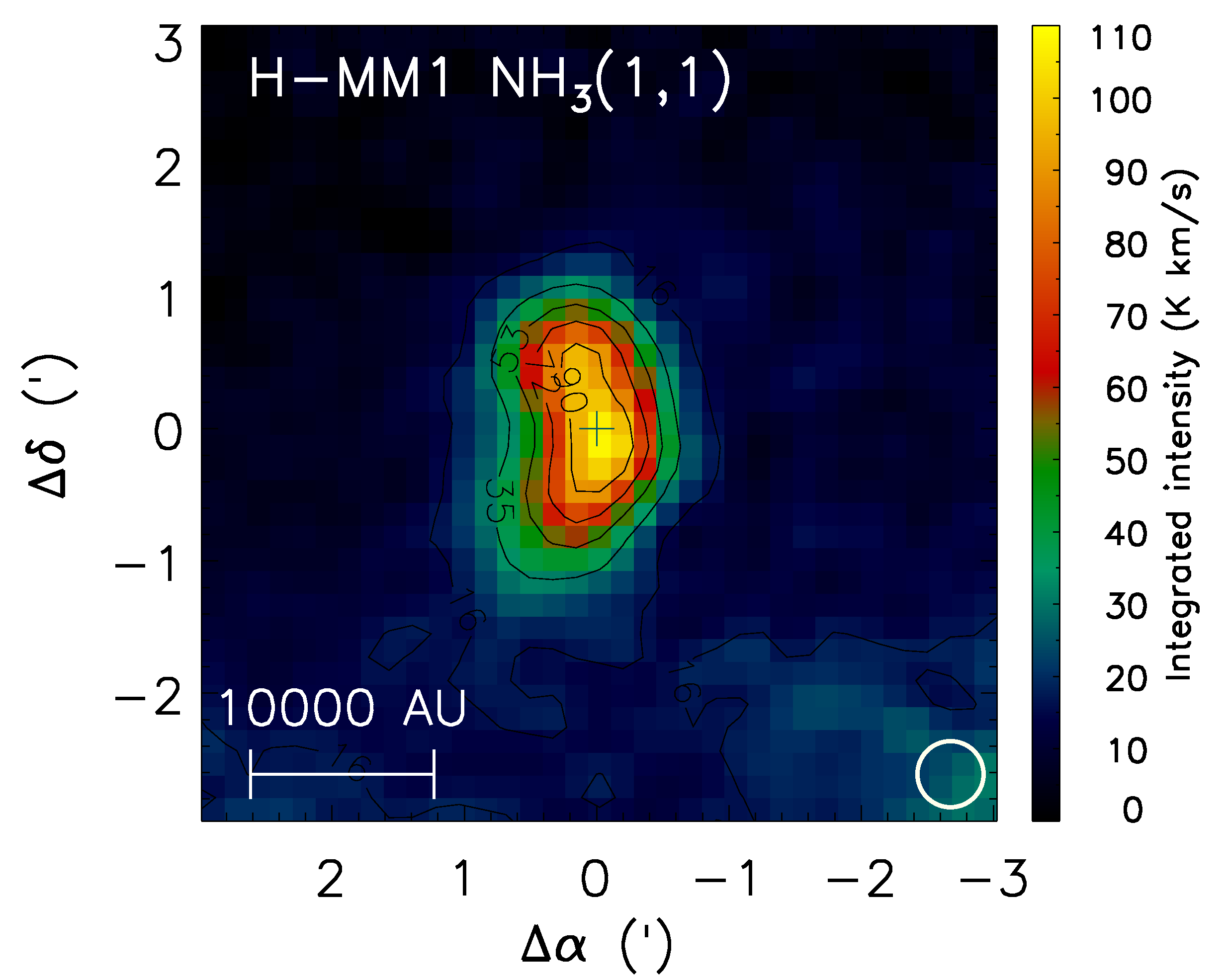}
\caption{Integrated $\ammo(1,1)$ intensity ($T_{\rm MB}$) map of H-MM1
  observed with the GBT. The intensity unit is K\,km\,s$^{-1}$, and the
  colour scale is given on the right. The $32\arcsec$ beam size of the
  GBT at 23.7 GHz is indicated in the bottom right. The APEX and IRAM spectra
 were taken towards the position indicated with a plus sign.}
\label{figure:nh311_map}
\end{figure}

The observations were carried out using the 7-beam K-Band Focal Plane
Array (KFPA) at the GBT, with the Versatile GBT Astronomical
Spectrometer (VEGAS) backend, as part of the Gould Belt Ammonia Survey
(GBT15A-430, PIs: Friesen \& Pineda).  VEGAS was configured in Mode 20
which uses 8 separate spectral windows per KFPA beam, each with a
bandwidth of 23.44 MHz and 4096 spectral channels. The spectral
resolution is 5.7 kHz, corresponding to $\sim0.07$~km\,s$^{-1}$.
Observations were performed using in-band frequency switching with a
frequency throw of 4.11 MHz.  Here we use the $\ammo(1,1)$ and $(2,2)$
line maps of a $6\arcmin\times6\arcmin$ region centered on the column
density peak of H-MM1. The integrated $\ammo(1,1)$ intensity map of
this region is shown in Fig.~\ref{figure:nh311_map}.

These observations were part of a much larger area map of the entire
L1688 region during the 15A semester, which will be presented by
Friesen \& Pineda et al. (in prep), and carried out on
$10\arcmin\times10\arcmin$ boxes scanned in right ascension with rows
separated by $13\arcsec$ in declination.  The scanning rate was
$6.2\arcsec$\,s$^{-1}$, with a data dump every 1.044 s.  A fast
frequency switching rate of 0.348 s was used, which results in an rms
of 0.1 K (on the $T_{\rm MB}$ scale).

The data is calibrated using the GBT KFPA data reduction pipeline
\citep{2011ASPC..442..127M}. The data were
calibrated to the $T_{\rm MB}$ scale using the gain factors for each beam
calibration derived from the Moon observations.  The final cubes are
created by a custom made gridder using a tapered Bessel function for
the convolution following \cite{2007A&A...474..679M}.
The full calibration and imaging pipeline is available to the
community at \url{https://github.com/GBTAmmoniaSurvey/GAS}.

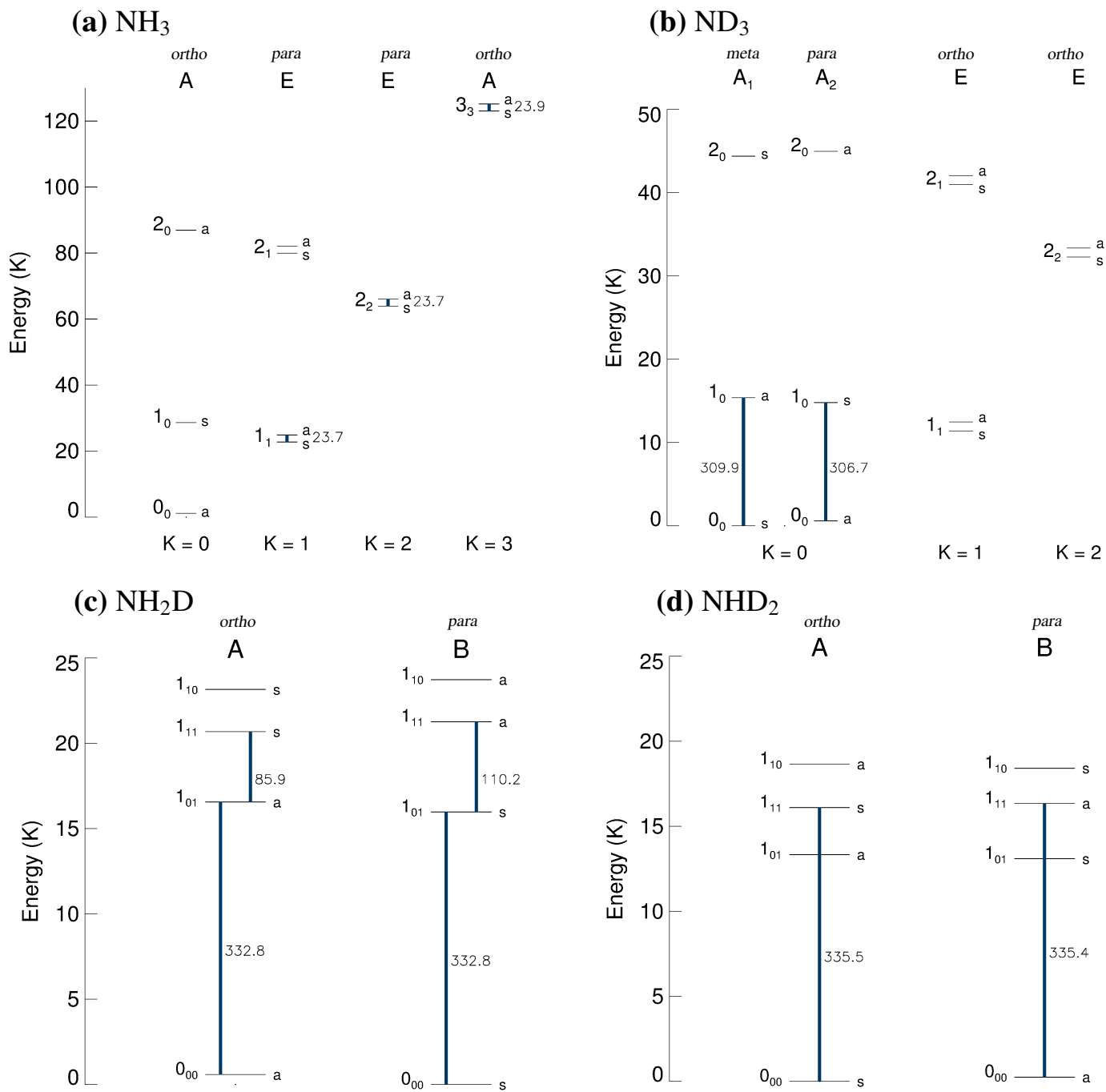
\begin{figure*}[ht]
\unitlength=1mm
\begin{picture}(180,177)(0,0)
\put(-5,86){
\begin{picture}(0,0) 
\includegraphics[width=9.3cm]{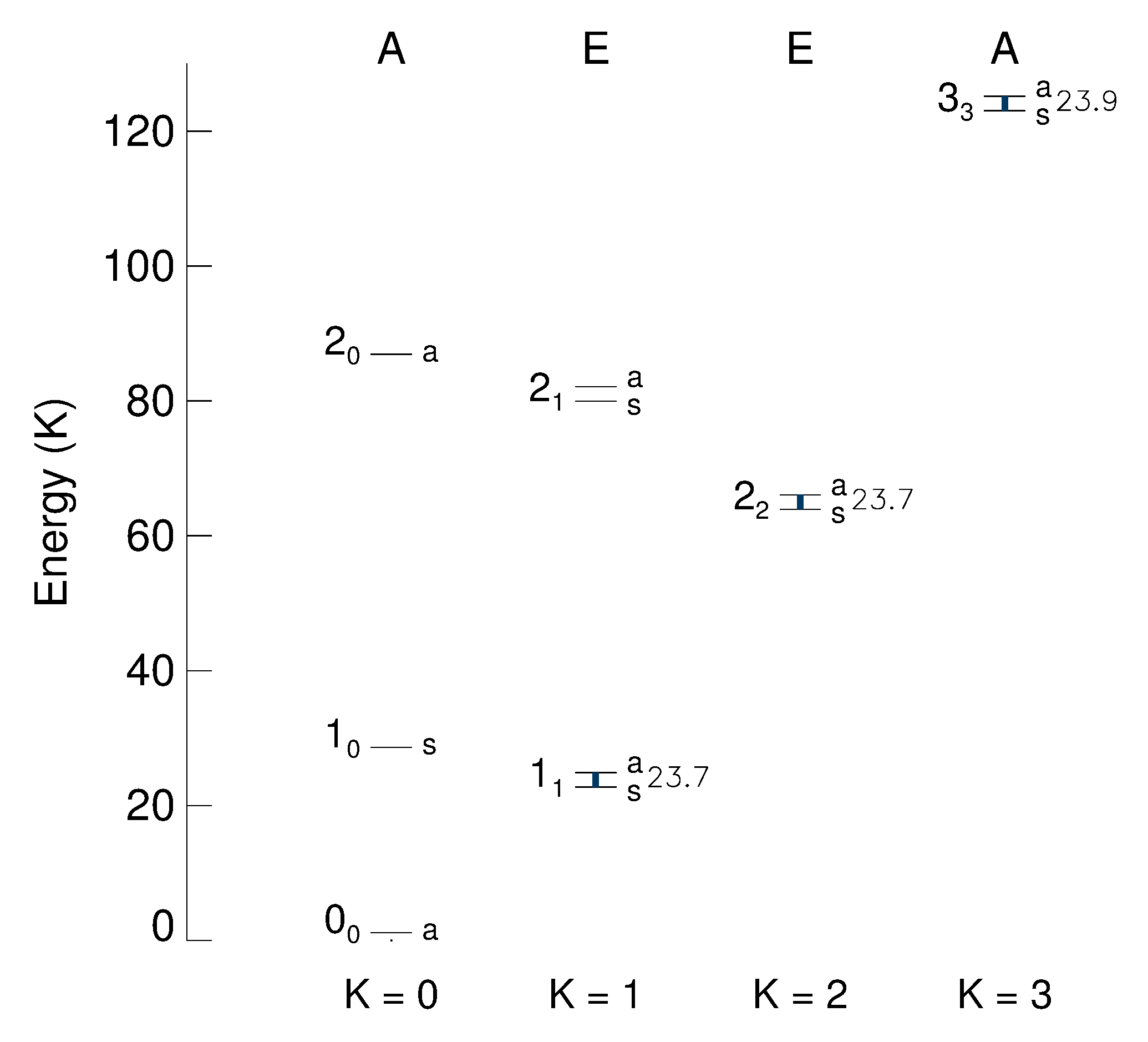}
\end{picture}}
\put(94,85){
\begin{picture}(0,0) 
\includegraphics[width=8.8cm]{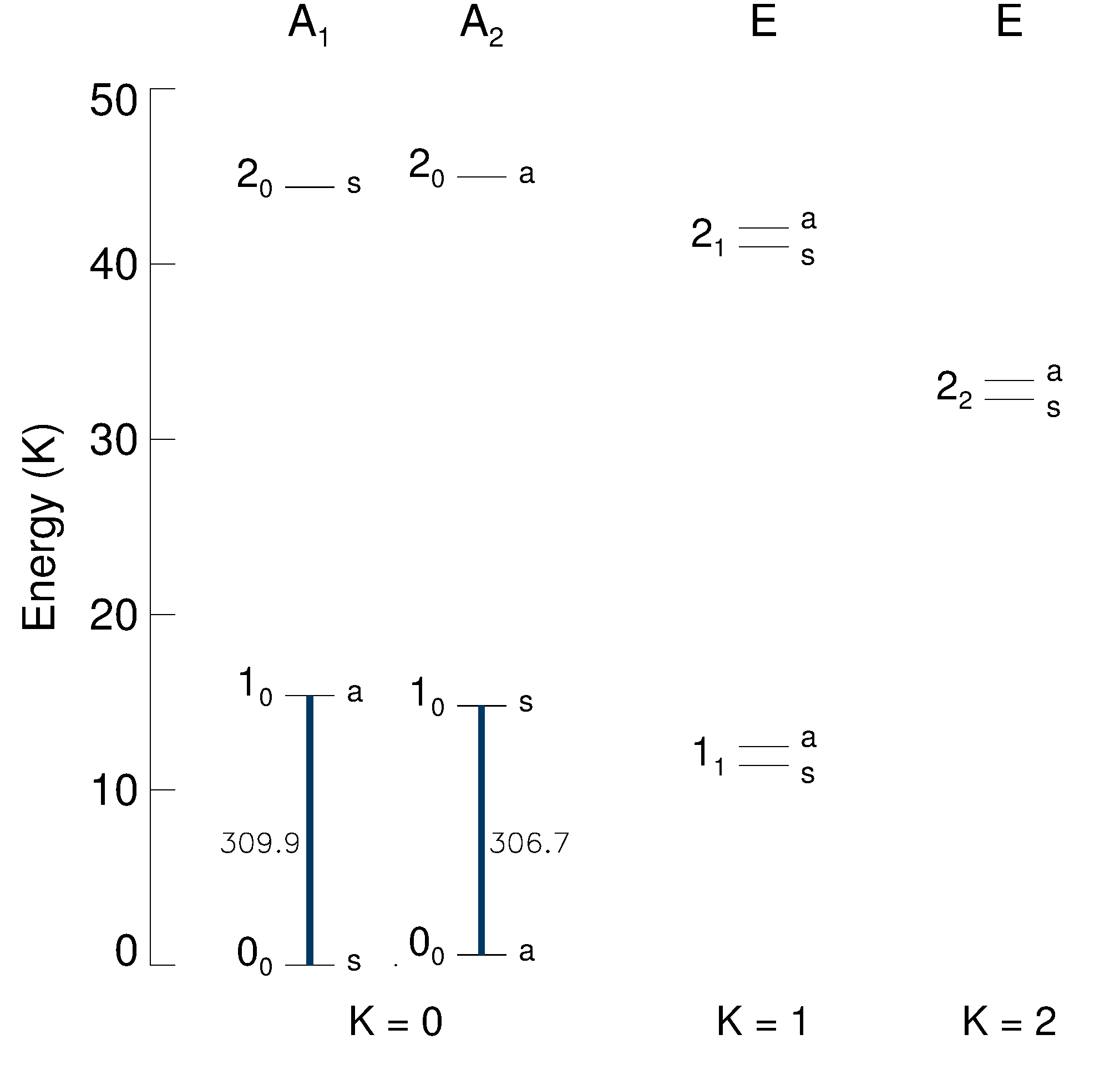}
\end{picture}}
\put(-2,-4){
\begin{picture}(0,0) 
\includegraphics[width=8.9cm]{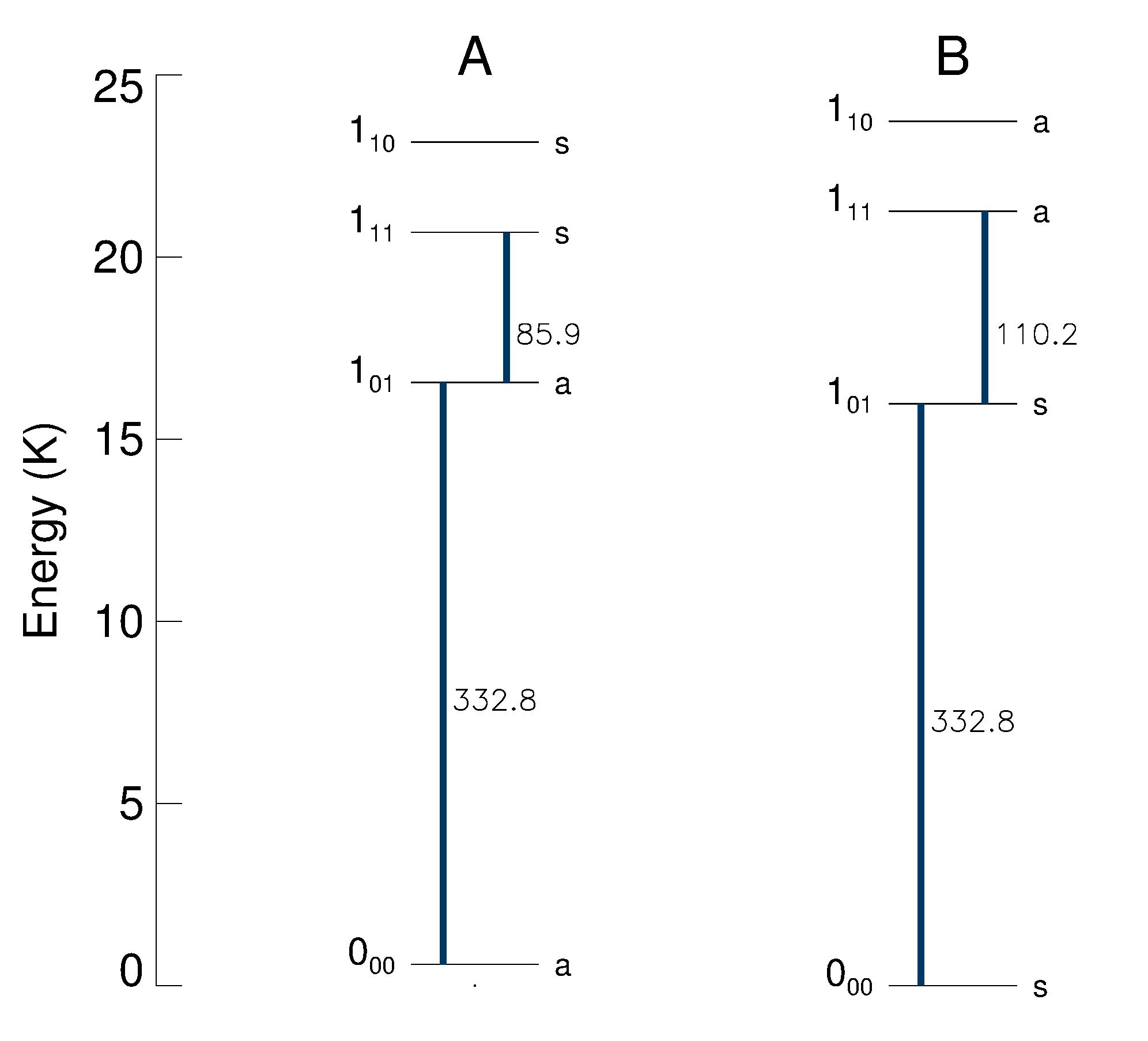}
\end{picture}}
\put(94,-4){
\begin{picture}(0,0) 
\includegraphics[width=8.8cm]{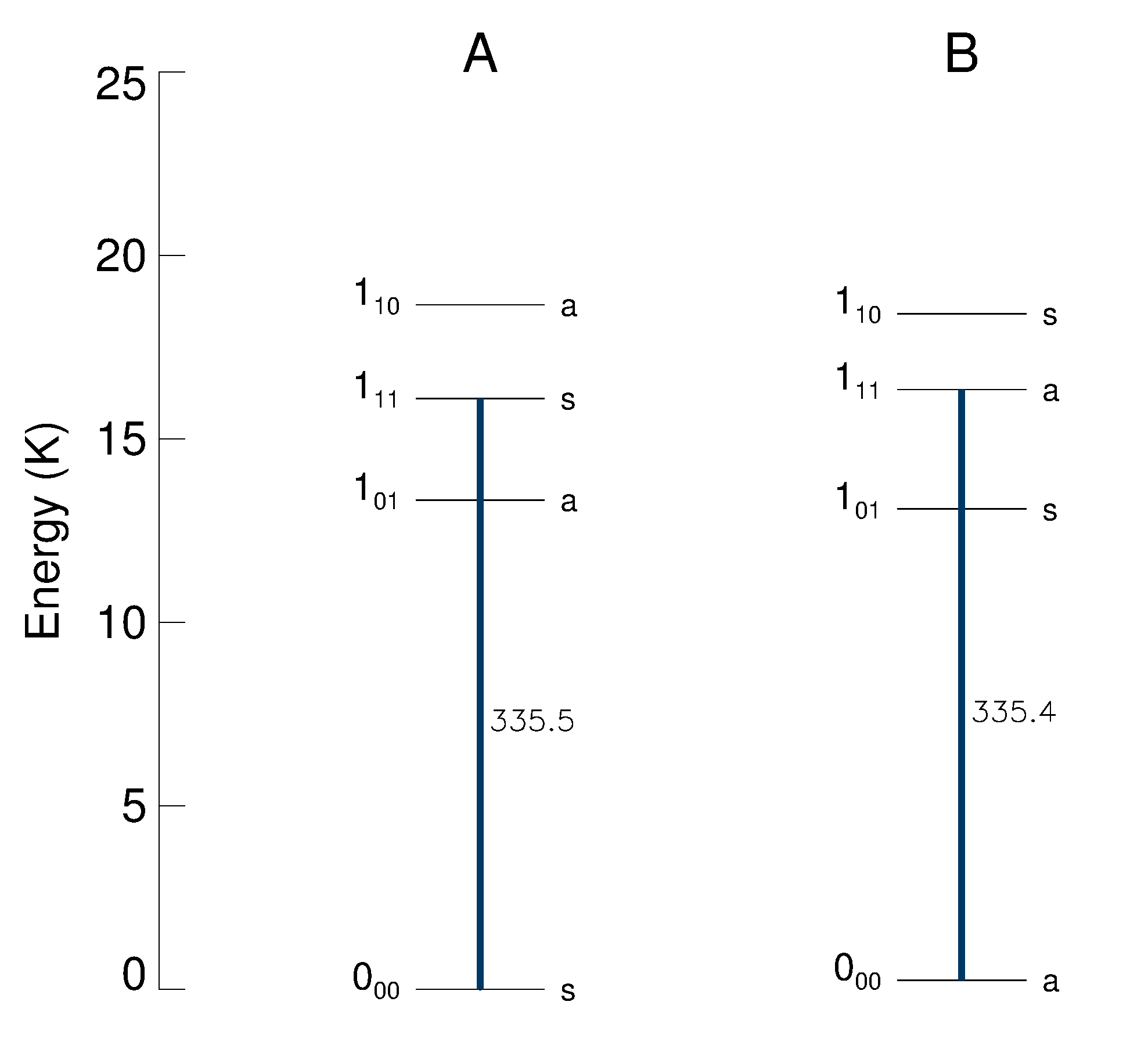}
\end{picture}}
\put(17,176){\makebox(0,0){\Large{\bf (a)} $\ammo$}}
\put(112,176){\makebox(0,0){\Large{\bf (b)} $\ndthree$}}
\put(19,80){\makebox(0,0){\Large{\bf (c)} $\dammo$}}
\put(115,80){\makebox(0,0){\Large{\bf (d)} $\ddammo$}}

\put(28,171){\makebox(0,0){\tiny {\sl ortho}}}
\put(44,170.5){\makebox(0,0){\tiny {\sl para}}}
\put(62,170.5){\makebox(0,0){\tiny {\sl para}}}
\put(78,171){\makebox(0,0){\tiny {\sl ortho}}}

\put(119,171){\makebox(0,0){\tiny {\sl meta}}}
\put(132,170.5){\makebox(0,0){\tiny {\sl para}}}
\put(154,171){\makebox(0,0){\tiny {\sl ortho}}}
\put(172,171){\makebox(0,0){\tiny {\sl ortho}}}

\put(36,77){\makebox(0,0){\tiny {\sl ortho}}}
\put(73,76.5){\makebox(0,0){\tiny {\sl para}}}
\put(132,77){\makebox(0,0){\tiny {\sl ortho}}}
\put(169,76.5){\makebox(0,0){\tiny {\sl para}}}

\end{picture}
\caption{Energies of the lowest rotational levels of $\ammo$ ({\bf
    a}), $\ndthree$ ({\bf b}), $\dammo$ ({\bf c}), and $\ddammo$ ({\bf
    d}). The nuclear spin symmetries and their ``para'', ``meta'', and
  ``ortho'' appellations are indicated. The ground-state
  rotation-inversion transition $1_0^{\rm s}-0_0^{\rm a}$ of {\sl
    ortho}-$\ammo$ at 572.5 GHz is only observable from space. The
  splitting between the inversion doublets of $\ammo$ has been
  exaggerated for clarity.}
\label{figure:energy_levels}
\end{figure*}

The observed ammonia transitions are indicated in the energy level diagram
shown in Fig.~\ref{figure:energy_levels}.

\subsection{APEX observations}

The centre position of H-MM1 was observed using the upgraded version
of the First Light APEX Submillimeter Heterodyne instrument
\citep[FLASH;][]{2006A&A...454L..21H} on APEX
\citep{2006A&A...454L..13G}. This instrument, FLASH$^+$
\citep{2014ITTST...4..588K}, operates simultaneously in the 345 GHz
and the 460 GHz atmospheric windows, and it can record two 4 GHz wide
sidebands separated by 12 GHz in both windows, i.e., altogether
$4\times4$ GHz frequency bands.  The receivers were connected to MPIfR
Fast Fourier Transform Spectrometers (XFTTS,
\citealt{2012A&A...542L...3K}) with spectral resolutions of $\sim
0.03$ and $\sim 0.05$ km\,s$^{-1}$ at 345 and 460 GHz, respectively.

The sky subtraction was done by position switching, using an absolute
reference position (R.A. 16$^{\rm h}$28$^{\rm m}$32$^{s}$,
Dec. $-24\degr31\arcmin00\arcsec$, J2000) which, judging from Herschel
far-infrared maps is void of dense gas. In the lower frequency window
we used two frequency settings which covered 1) the ground-state lines
of {\sl para}- and {\sl meta}-$\ndthree$ (hereafter p$\ndthree$ and
m$\ndthree$) at 306.7 and 309.9 GHz, and 2) the ground-state lines of
{\sl ortho}- and {\sl para}-$\dammo$ (o$\dammo$ and p$\dammo$) at
332.8 GHz, and the ground-state lines of {\sl ortho}- and {\sl
  para}-$\ddammo$ (o$\ddammo$ and p$\ddammo$) at 335.5 GHz. The first
tuning covered also the $\ddiaz(4-3)$ line at 308.4 GHz.  The
observations in the 460 GHz window were aimed at the $\ddiaz(6-5)$ and
$\diaz(5-4)$ lines. Because these two lines could be measured
simultaneously the tuning of the 460 GHz receiver was kept constant
during the whole observing run. 

A list of transitions covered is given in Table~\ref{table:obstrans},
mentioning only the most significant to the present study. Here we
give the centre frequencies of transitions, upper state energies, and Einstein
coefficients for spontaneous emission. These parameters are obtained
from the Cologne Database for Molecular Spectroscopy,
CDMS\footnote{\url{www.astro.uni-koeln.de/cdms/catalog}}.  The
observed transitions of $\ammo$, $\dammo$, $\ddammo$, and $\ndthree$
are indicated in the energy level diagrams in
Figs.~\ref{figure:energy_levels}.

The APEX beamsize (FWHM) is $\sim 20\arcsec$ at $310-330$ GHz, and
$\sim 14\arcsec$ at 465 GHz. The main beam efficiency, $\eta_{\rm
  MB}$, is 0.73 at $310-330$ GHz, and 0.6 at 465 GHz
\citep{2006A&A...454L..13G}.  The observations were carried out
between 29 and 31 May, 2015. The total observing time was 11.4
hours. The weather conditions were stable and fairly good (PWV 0.7-1.2
mm). The absolute calibration, pointing, and focus were checked by
observing Saturn.  The system temperatures were in the following
ranges: $180-190$ K (310 GHz), $230-250$ K (335 GHz), and $570-600$ K
(465 GHz). The resulting RMS noise levels at the mentioned frequencies
at a velocity resolution of 0.1 km\,s$^{-1}$, were 18, 27, and 33 mK,
respectively, on the $T_{\rm MB}^*$ scale.

\begin{table}
\caption{Observed transitions.}
\begin{tabular}{lllll}
\hline \hline
\multicolumn{2}{c}{transition} & frequency & $E_{\rm upper}$ & $A_{\rm ul} $ \\
\multicolumn{2}{c}{}     & (MHz)     &  (K)         & (s$^{-1}$) \\ \hline
\multicolumn{5}{c}{ } \\
\multicolumn{5}{c}{GBT 100-m} \\
\multicolumn{5}{c}{ } \\
p$\ammo$&$(1_1^{\rm a}-1_1^{\rm s})$ & 23694.4955 & 24.4 & $1.68\,10^{-7}$ \\
p$\ammo$&$(2_2^{\rm a}-2_2^{\rm s})$ & 23722.6336 & 65.6 & $2.24\,10^{-7}$  \\
o$\ammo$&$(3_3^{\rm a}-3_3^{\rm s})$ & 23870.1296 & 124.7 &$2.57\,10^{-7}$  \\
\multicolumn{5}{c}{ } \\
\multicolumn{5}{c}{APEX 12-m} \\
\multicolumn{5}{c}{ } \\
o$\dammo$&$(1_{01}^{\rm a}-0_{00}^{\rm a})$ & 332781.890 &  16.6 & $8.14\,10^{-6}$ \\
p$\dammo$&$(1_{01}^{\rm s}-0_{00}^{\rm s})$ & 332822.510 &  16.0 & $7.60\,10^{-6}$ \\
o$\ddammo$&$(1_{11}^{\rm s}-0_{00}^{\rm s})$& 335513.793 &  16.1 & $1.29\,10^{-5}$ \\ 
p$\ddammo$&$(1_{11}^{\rm a}-0_{00}^{\rm a})$& 335446.321 &  16.3 & $1.47\,10^{-5}$ \\ 
m$\ndthree$&$(1_0^{\rm a}-0_0^{\rm s})$  & 309909.490 &  14.9 & $2.59\,10^{-4}$ \\ 
p$\ndthree$&$(1_0^{\rm s}-0_0^{\rm a})$  & 306736.710 &  14.8 & $2.51\,10^{-4}$ \\
$\ddiaz$&$(4-3)$           & 308422.267 &  37.0 &  $1.75\,10^{-3}$ \\
$\ddiaz$&$(6-5)$           & 462603.852 &  77.7 &  $6.15\,10^{-3}$ \\   
$\diaz$&$(5-4)$            & 465824.777 &  67.1 &  $6.18\,10^{-3}$ \\
\multicolumn{5}{c}{ } \\
\multicolumn{5}{c}{IRAM 30-m} \\
\multicolumn{5}{c}{ } \\
o$\dammo$&$(1_{11}^{\rm s}-1_{01}^{\rm a})$ &  85926.278  &  20.7 & $7.84\,10^{-6}$ \\
p$\dammo$&$(1_{11}^{\rm a}-1_{01}^{\rm s})$ & 110153.594  &  21.3 & $1.65\,10^{-5}$ \\
$\ddiaz$&$(2-1)$           & 154217.011 &  11.1 &  $1.97\,10^{-4}$ \\
\end{tabular}
\label{table:obstrans}
\end{table}

\subsection{IRAM observations}

The column density peak of H-MM1 was observed with the IRAM 30~meter
telescope on July 5, 2015 in acceptable weather conditions (PWV
8--10~mm). Pointing and focus were checked towards QSO 1253-055.  The
following transitions were observed: o$\dammo(1_{11}-1_{01})$ at
85.9~GHz, p$\dammo(1_{11}-1_{01})$ at 110.2~GHz, and $\ddiaz(2-1)$ at
154.2~GHz.  The measurements were obtained with the EMIR 090 and 150
receivers\footnote{\url{www.iram.es/IRAMES/mainWiki/EmirforAstronomers}} and
the VESPA autocorrelator with spectral resolution of 20~kHz; the
corresponding velocity resolutions were 0.04--0.07~km~s$^{-1}$. The beam
sizes were $29\arcsec$, $23\arcsec$, and $16\arcsec$ for
o$\dammo(1_{11}-1_{01})$, p$\dammo(1_{11}-1_{01})$, and $\diaz(2-1)$,
respectively. The system temperatures ranged from 166 to 310~K
depending on the frequency (see Table~\ref{observations} for the
details).  The spectra were taken in the position switching mode,
using an off position 400$^{\prime\prime}$ East of the target. The
integration time for each line was between 15 and 22~minutes which
resulted in RMS noise levels of 0.08--0.14~K on the $T_{\rm MB}$
scale. The intensity scale was converted to the main-beam temperature
scale using the beam efficiency values given in the IRAM 30~m report
on a beam pattern~\citep{Kramer2013}; see Table~\ref{observations} for
details.

\begin{table*}
\caption{Observational parameters.}\label{observations}
\begin{tabular}{llcccccccc}
\hline \hline
Molecule&Transition&Frequency&HPBW&$\eta_{\rm fss}$& $\eta_{\rm MB}$&
$\Delta \upsilon^a$&$T_{\rm sys}$&time&RMS$^b$\\
 & &(GHz)&($\arcsec$)& & &(km~s$^{-1}$)&(K)&(min)&(K)\\
\hline
\multicolumn{10}{c}{} \\
\multicolumn{10}{c}{GBT} \\
\multicolumn{10}{c}{} \\
p$\ammo$&$(1_{1}^{\rm a}-1_{1}^{\rm s})$&23.7&32& 0.95 &0.91 &0.072& 45 & 94 &0.103\\
p$\ammo$&$(2_{2}^{\rm a}-2_{2}^{\rm s})$&23.7&32& 0.95 &0.91 &0.072& 45 & 94 &0.083\\
o$\ammo$&$(3_{3}^{\rm a}-3_{3}^{\rm s})$&23.9&32& 0.95 &0.91 &0.072& 45 & 94 &0.096\\
\multicolumn{10}{c}{} \\
\multicolumn{10}{c}{APEX} \\
\multicolumn{10}{c}{} \\
o$\dammo$&$(1_{01}^{\rm a}-0_{00}^{\rm a})$&332.8&19&0.97&0.73&0.040&263&86&0.027\\
p$\dammo$&$(1_{01}^{\rm s}-0_{00}^{\rm s})$&332.8&19&0.97&0.73&0.040&263&86&0.027 \\
o$\ddammo$&$(1_{11}^{\rm s}-0_{00}^{\rm s})$&335.5&19&0.97&0.73&0.040&227&86&0.021\\
p$\ddammo$&$(1_{11}^{\rm a}-0_{00}^{\rm a})$&335.4&19&0.97&0.73&0.040&227&86&0.021 \\
m$\ndthree$&$(1_0^{\rm a}-0_0^{\rm s})$&309.9&20&0.97&0.73&0.043&195&75&0.019\\
p$\ndthree$&$(1_0^{\rm s}-0_0^{\rm a})$&306.7&20&0.97&0.73&0.043&178&75&0.019\\
$\ddiaz$&$(4-3)$&308.4&20&0.97&0.73&0.043&178&75&0.019\\
$\ddiaz$&$(6-5)$&462.6&14&0.95&0.60&0.057&594&97&0.043\\
$\diaz$&$(5-4)$&465.8&14&0.97&0.60&0.057&557&108&0.038\\
\multicolumn{10}{c}{} \\
\multicolumn{10}{c}{IRAM} \\
\multicolumn{10}{c}{} \\
o$\dammo$&$(1_{11}^{\rm s}-1_{01}^{\rm a})$&85.9&29&0.95&0.81&0.068&166&15&0.080\\
p$\dammo$&$(1_{11}^{\rm a}-1_{01}^{\rm s})$&110.2&23&0.94&0.79&0.053&249&22&0.098\\
$\ddiaz$&$(2-1)$&154.2&16&0.93&0.72&0.038&310&22&0.137\\ \hline
\end{tabular}

$^a$ Spectral resolution ($=$ equivalent noise bandwidth, ENBW). $^b$ On the $T_{\rm MB}$ scale at the original spectral resolution.

\end{table*}

\section{LTE Analysis of the observed spectra}

\label{sec:LTE_analysis}

In this section we present the observed spectra and the results of the
standard hyperfine stucture fitting to the detected lines. The method
assumes line-of-sight homogeneity and that the populations of
hyperfine states in a certain rotational manifold are proportional to
their statistical weights, according to the assumption of local
thermodynamic equilibrium, LTE. The total column density estimates
rely furthermore on the assumption that the excitation temperature,
$T_{\rm ex}$, is constant for all rotational transitions of a
molecule.

\begin{figure}
\includegraphics[width=8cm]{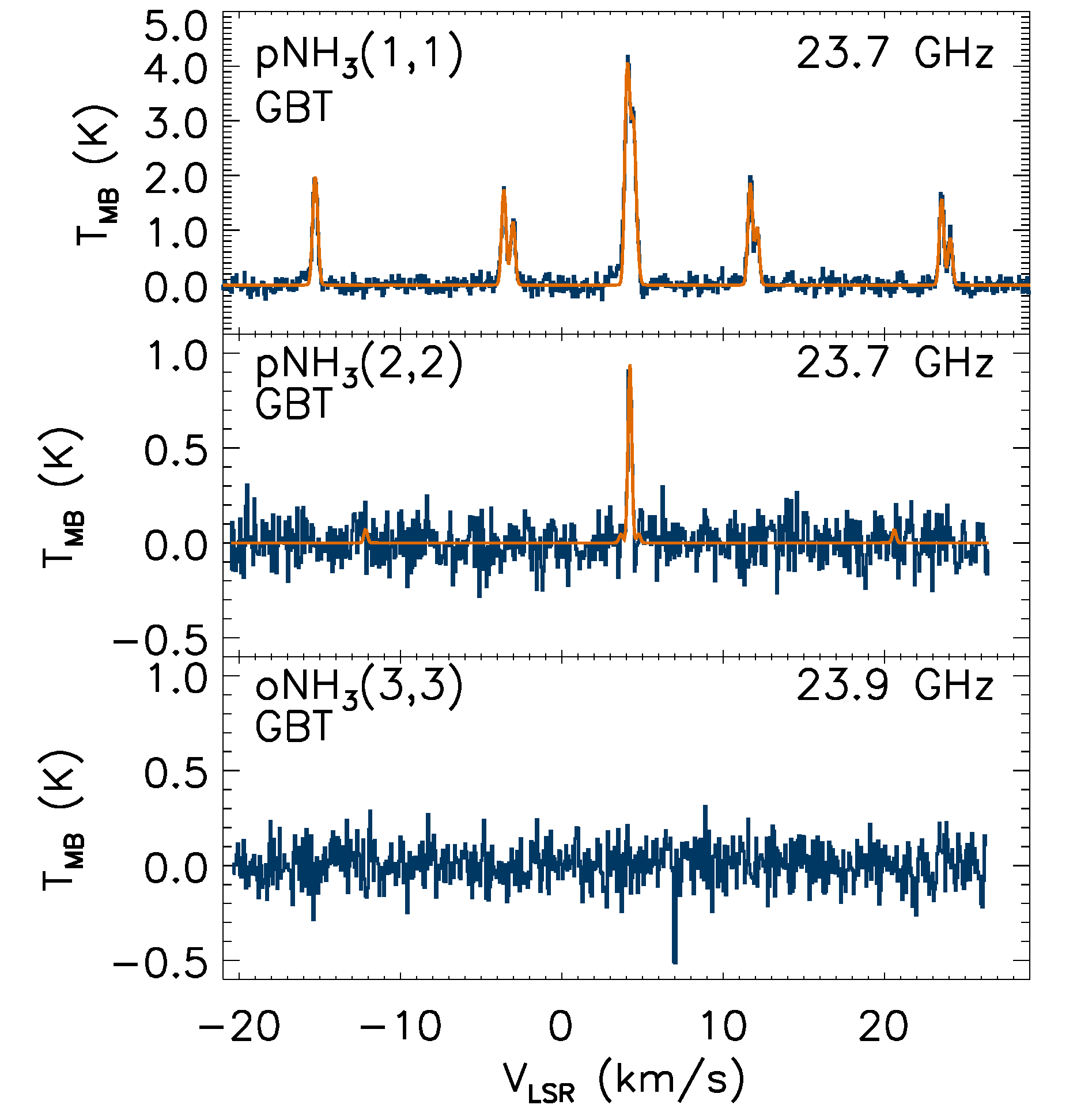}
\caption{$\ammo(1,1)$, $(2,2)$, and $(3,3)$ inversion line spectra
  observed with the GBT towards the centre of H-MM1. The spectra are
  on the $T_{\rm MB}$ scale. Fits to the $(1,1)$ and $(2,2)$ hyperfine
  structures are indicated with orange curves. The origin of the
  absorption feature seen in the $(3,3)$ spectrum (at 23869.6 MHz) is
  unkown to us.}
\label{figure:ammo}
\end{figure}

\begin{figure}
\unitlength=1mm
\begin{picture}(90,80)(0,0)
\put(-3,0){
\begin{picture}(0,0) 
\includegraphics[width=9.5cm]{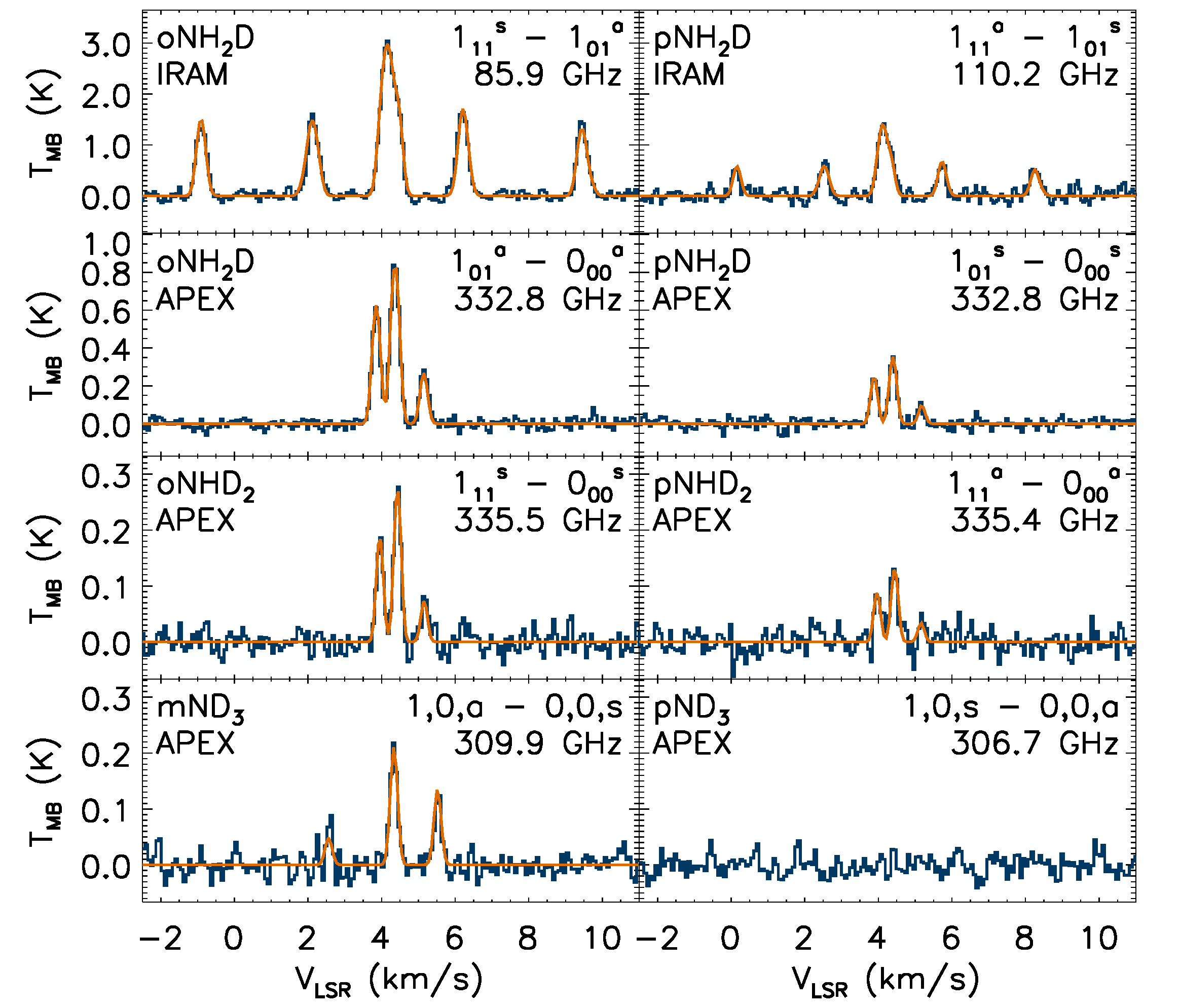}
\end{picture}}
\end{picture}
\caption{Deuterated ammonia spectra observed at IRAM and APEX. The
  spectra are presented on the $T_{\rm MB}^*$ scale. The APEX spectra
  are Hanning smoothed to a resolution of about 0.07 km\,s$^{-1}$
  which corresponds to the spectral resolution of the spectra from
  GBT and IRAM. The orange curves are Gaussian fits to the hyperfine structure.}
\label{figure:dammo}
\end{figure}

\begin{figure}
\unitlength=1mm
\begin{picture}(90,100)(0,0)
\put(-5,0){
\begin{picture}(0,0) 
\includegraphics[width=9.0cm]{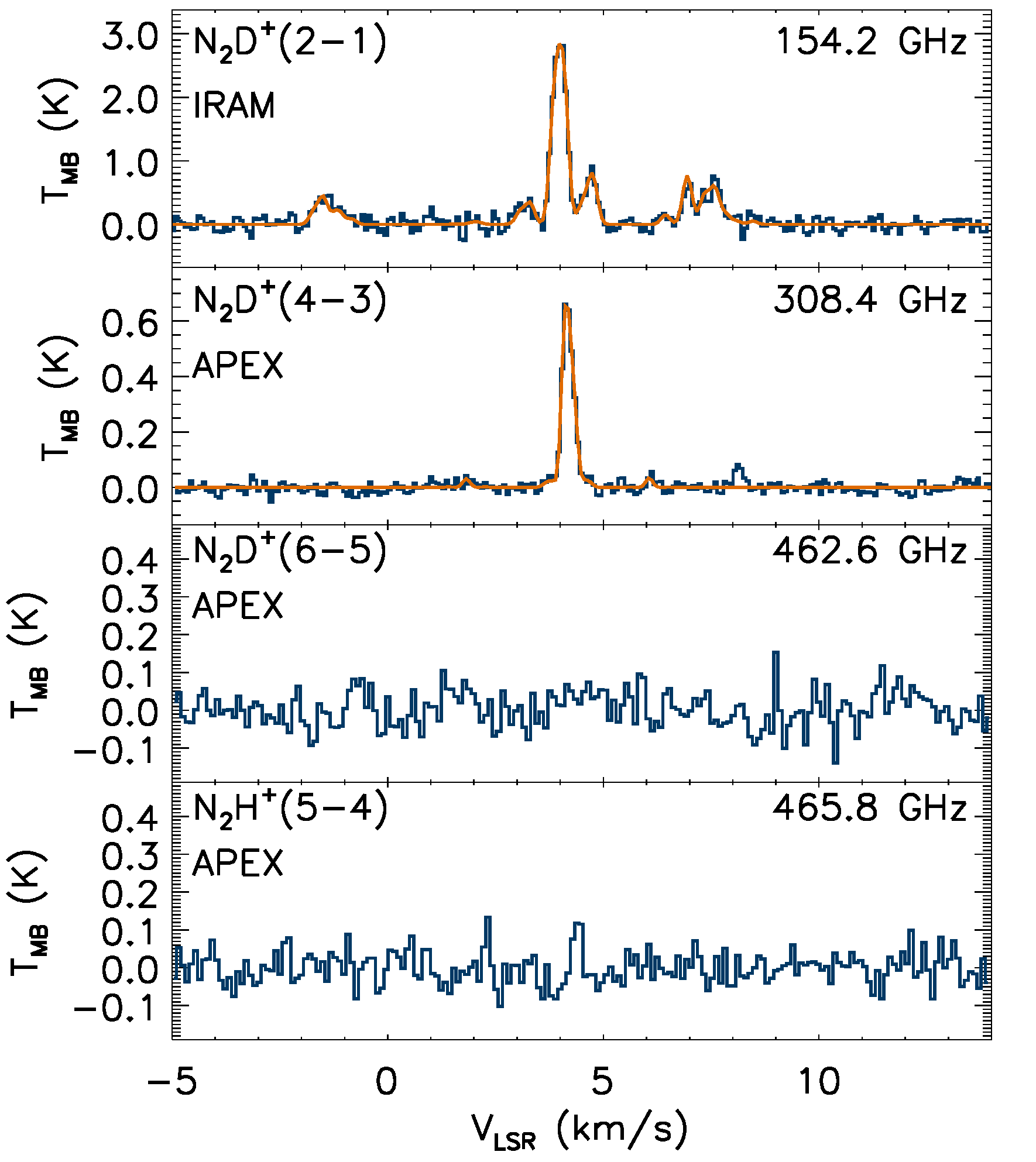}
\end{picture}}
\end{picture}
\caption{$\ddiaz$ and $\diaz$ spectra observed at IRAM and APEX. The
  APEX spectra are Hanning smoothed. The velocity resolution is about
  0.07 km\,s$^{-1}$ for the $\ddiaz(2-1)$ and $\ddiaz(4-3)$ spectra
  and about 0.1 km\,s$^{-1}$ for the higher frequency
  spectra. Gaussian fits to the hyperfine components of the
  $\ddiaz(2-1)$ and $\ddiaz(4-3)$ lines are shown as orange curves.}
\label{figure:diaz}
\end{figure}

The observed spectra are shown in Figs.~\ref{figure:ammo} ($\ammo$),
\ref{figure:dammo} ($\dammo$, $\ddammo$, and $\ndthree$), and
\ref{figure:diaz} ($\ddiaz$ and $\diaz$). The spectra were reduced
using the GILDAS software package\footnote{Grenoble Image and Line
  Data Analysis Software package has been developed by IRAM-Grenoble,
  see \url{www.iram.fr/IRAMFR/GILDAS}}.  All the observed transitions have
hyperfine structure. The detected lines were analysed using the HFS
method implemented in the CLASS software (part of GILDAS), or fitting
routines written in the IDL language.

The o$\ammo(3,3)$ and $\ddiaz(6-5)$ lines were not detected, while
$\diaz(5-4)$ shows perhaps a weak line with an integrated intensity of
$\sim 30$ mK\,km\,s$^{-1}$. The upper limits for the intensities of
these lines are 0.1 K, 0.05 K and 0.1 K, respectively, on the $T_{\rm
  MB}$ scale. For the $\ddiaz(6-5)$ and $\diaz(5-4)$ spectra we have
only used scans (20 s integrations) with smooth baselines, showing no
visible disturbances, which means about half of the measurements.

The hyperfine patterns of the detected lines are dominated by the
splitting caused by the electric quadrupole moment of the $^{14}$N
nucleus.  In the $\ammo$ inversion lines, the structure owing to the
magnetic moments of the N and H nuclei can be partially resolved
(\citealt{1967PhRv..156...83K}; \citealt{1983ARA&A..21..239H}).
Furthermore, for the $\dammo$ lines we use the hyperfine patterns
calculated by \cite{2016A&A...586L...4D}, including the effects of
both N and D nuclei. As shown recently by \cite{2016A&A...586L...4D},
quadrupole coupling of the D nucleus broadens substantially the
observed $1_{11}-1_{01}$ lines of $\dammo$ at 86 and 110 GHz, and
disregard of this effect results in overestimates of the line widths
by $\sim 50\%$ for a cold cloud. For the ground-state lines at 333
GHz, and for any lines at higher frequencies, the effect of the
hyperfine structure owing to D is, however, negligible compared with
the Doppler broadening.

For other molecules observed here we use line lists which only take
the splitting due to N into account.  The data are from
\cite{2006A&A...449..855C} ($\ddammo$ and $\ndthree$), and from
\cite{1995ApJ...455L..77C}, \cite{2004A&A...413.1177D}, and
\cite{2009A&A...494..719P} ($\diaz$ and $\ddiaz$). The effect of other
interactions is likely to be small compared with the thermal
broadening (\citealt{1969JChPh..49.5523K};
\citealt{2004A&A...413.1177D}). In the hyperfine fitting, we have
assumed that the frequencies listed in Table~\ref{table:obstrans},
which are adopted from the CDMS, represent weighted averages of the
hyperfine components.

The results of Gaussian fits to the hyperfine structure are presented
in Table~\ref{table:hffits}. This table contains the peak main-beam
brightness temperatures, $T_{\rm MB}$, the radial velocities, $V_{\rm
  LSR}$, the line widths, $\Delta \upsilon$, and the sums of the peak
optical thicknesses, $\tau_{\rm sum}$, of the hyperfine
components. The hyperfine fit also gives an estimate for the 
$T_{\rm ex}$ of the transition, provided that
the spectrum is on the brightness temperature scale. We have
approximated this by the $T_{\rm MB}$ scale, which is equivalent to
assuming that the source fills the telescope beam uniformly. Finally,
the two last columns of Table~\ref{table:hffits} give the total column
densities, $N_{\rm total}$, of the molecules, and the fractional
abundances, $X$, using the $\htwo$ column density derived from
Herschel observations. The derived total optical thicknesses of the
detected lines range from $\sim 1$ to $\sim 5$. This means that the
satellites are mainly optically thin, and we are in the regime where
the integrated intensity is proportional to the column density of the
molecule. 

\begin{table*}
  \caption{Hyperfine fit results, column densities, and 
    the fractional abundances relative to $\htwo$.}
\begin{tabular}{ll|c|c|c|c|c|c|c} 
\hline \hline
\multicolumn{2}{c|}{transition} & $T_{\rm MB}$ (K) & $V_{\rm LSR}$ (km\,s$^{-1}$) &
$\Delta \upsilon$ (km\,s$^{-1}$) & $\tau_{\rm sum}$ &
$T_{\rm ex}$ (K) &
$N_{\rm total}$ (cm$^{-2}$) & X \\ \hline
p$\ammo$&$(1_{1}^{\rm a}-1_{1}^{\rm s})$&$4.18\pm0.10$&$4.222\pm0.002$&$0.297\pm0.005$&
               $4.8\pm0.3$&$8.2\pm0.2$&$(1.7\pm0.2)\,10^{14}$&$3\,10^{-9}$\\
p$\ammo$&$(2_{2}^{\rm a}-2_{2}^{\rm s})$&$0.91\pm0.08$&$4.220\pm0.008$&$0.256\pm0.020$&
               $0.22\pm0.09$&$9.2\pm0.2$& & \\
o$\dammo$&$(1_{11} - 1_{01})$&$3.04\pm0.08$&$4.283\pm0.002$&$0.235\pm0.006$&$5.1\pm0.3$&
               $7.0\pm0.2$&$(1.1\pm0.1)\,10^{14}$&$2\,10^{-9}$\\
o$\dammo$&$(1_{01} - 0_{00})$& 
$0.84\pm0.03$ & $4.327\pm0.003$ & $0.223\pm0.006$ & $2.4\pm0.4$ & 
$6.1\pm0.2$ & $(1.2\pm 0.3)\,10^{14}$ & \\

p$\dammo$&$(1_{11} - 1_{01})$& $1.42\pm0.11$ & $4.285\pm0.005$ & $0.222\pm0.016$ & $2.3\pm0.8$ &
$5.9\pm0.7$ & $(4.6\pm1.2)\,10^{13}$ &$8\,10^{-10}$ \\ 

p$\dammo$&$(1_{01} - 0_{00})$& 
$0.36\pm0.02$&  $4.247\pm0.005$&  $0.193\pm0.014$&  $1.5\pm0.8$& 
$5.1\pm0.6$& $(5.7\pm 4.2)\,10^{13}$ & \\

o$\ddammo$&$(1_{11} - 0_{00})$& 
$0.28\pm0.02$&  $4.239\pm0.005$&  $0.197\pm0.012$& $1.5\pm0.8$& 
 $4.7\pm0.4$& $(4.0\pm 3.2)\,10^{13}$ & $\sim7\,10^{-10}$ \\

p$\ddammo$&$(1_{11} - 0_{00})$&
$0.13\pm0.02$&  $4.261\pm0.010$&  $0.202\pm0.026$&  $\sim 1.3$ & 
$\sim 4.0$ & $\sim 2.7\,10^{13}$ & $\sim5\,10^{-10}$ \\

m$\ndthree$&$(1_0 - 0_0)$&
  $0.22\pm0.02$ & $4.336\pm0.006$ & $0.202\pm0.015$ & $1.1\pm0.5$ &
 $4.4\pm0.3$ & $(1.0\pm 0.5)\,10^{12}$ & $2\,10^{-11}$ \\

$\ddiaz$ & $(2-1)$ & $2.81\pm0.09$ & $4.197\pm0.003$ & $0.232\pm0.008$ & $3.5\pm0.3$ &
$7.4\pm0.2$ & $(3.4\pm0.5)\,10^{12}$ & $6\,10^{-11}$ \\

$\ddiaz$ & $(4-3)$ & $0.66\pm0.02$ & $4.273\pm0.003$ & $0.167\pm0.013$ & $2.4\pm0.7$ &
$5.2\pm0.2$ & $(1.7\pm0.7)\,10^{13}$ & \\

\end{tabular}
\label{table:hffits}
\end{table*}

While the accuracy of the LSR velocities and line widths from the
hyperfine fits is high, the optical thicknesses, and consequently the
column densities derived using this method have relatively large
uncertainties, except for p$\ammo$ and o$\dammo$, which are the brightest
lines.  The o$\dammo$ and p$\dammo$ column densities derived from the
lines observed with APEX and IRAM are consistent when the different
beam sizes (see Table 2) and the uncertainties owing to noise are
taken into account.  In contrast, the $\ddiaz$ column densities
derived from the $\ddiaz(2-1)$ (IRAM) and $\ddiaz(4-3)$ (APEX) spectra
differ by a factor of five. We consider the value obtained from the
$\ddiaz(2-1)$ spectrum more reliable because this line has several
resolved hyperfine components, and because this transition connects
rotational levels which are much more densely populated than $J=3$ and
$J=4$. Furthermore, the $\ddiaz$ column density obtained from
$\ddiaz(2-1)$ is consistent with that of derived from $\ddiaz(1-0)$ by
\cite{2016A&A...587A.118P} ($3.8\pm 0.9\,10^{12}$ cm$^{-2}$, beamsize
$32\arcsec$) towards a position lying $13\arcsec$ off from our centre
position. 

Assuming that {\sl ortho} and {\sl para} $\ammo$ have equal abundances
(which would correspond to their nuclear spin statistical weights), we
obtain a total ammonia column density of $N(\ammo) =
(3.3\pm0.4)\,10^{14}$ cm$^{-2}$.  The total column densities for
$\dammo$ and $\ddammo$ are $N(\dammo) = (1.5\pm0.1)\,10^{14}$
cm$^{-2}$, $N(\ddammo) \sim 6.7\,10^{13}$ cm$^{-2}$. Like in the case
of $\ammo$, we only have detected one spin modification of $\ndthree$.  Assuming
that the relative abundances of {\sl ortho}, {\sl meta}, and {\sl
  para} $\ndthree$ correspond to their nuclear spin statistical
weights, $16:10:1$, we get an estimate for total $\ndthree$ column
density with large error margins: $N(\ndthree)=(2.7\pm1.4)\,10^{12}$
cm$^{-2}$. These column density estimates imply the following
fractionation ratios: $\dammo/\ammo = 0.45\pm 0.09$, $\ddammo/\dammo
\sim 0.45$, and $\ndthree/\ddammo \sim 0.04$. The spin ratios can only
be estimated for $\dammo$ and $\ddammo$ for which we get o/p$\dammo =
3.0\pm1.1$ and o/p$\ddammo\sim1.5$. We see that both {\sl ortho}/{\sl
  para} ratios are close to their statistical values, 3 and 2,
respectively.

\begin{figure}
\includegraphics[width=9cm]{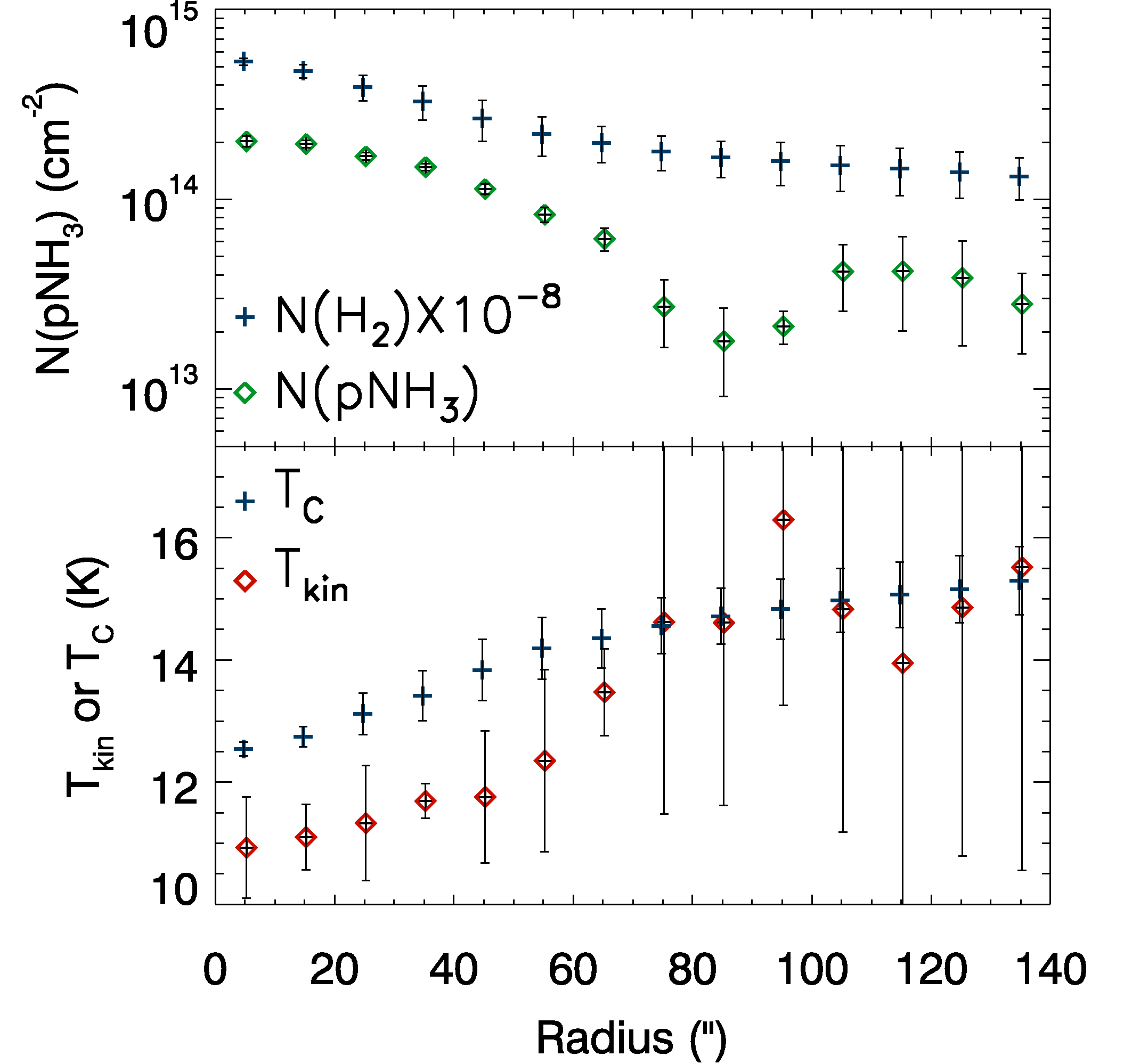}
\caption{Distributions of the $N(\ammo)$ and $N(\htwo)$ (top)
  and $T_{\rm kin}$ and $T_{\rm C}$ (bottom), as functions of the
  angular distance from the core centre. The $N(\ammo)$ and $T_{\rm
    kin}$ estimates are derived from the $\ammo(1,1)$ and $(2,2)$
  maps, whereas $N(\htwo)$ and $T_{\rm C}$ distributions are based on
  Hershel/SPIRE far-infrared continuum maps.}
\label{figure:tkin}
\end{figure}

We used the $\ammo(1,1)$ and $(2,2)$ maps from GBT to derive the
kinetic temperature, $T_{\rm kin}$, and the p$\ammo$ column density,
$N({\rm p}\ammo)$, distributions in the vicinity of the H-MM1.  The
standard analysis described in \cite{1983ARA&A..21..239H},
\cite{1983A&A...122..164W}, and \cite{1986A&A...157..207U} was
used. According to the modelling results of
\cite{2012A&A...538A.133J}, the ammonia spectra trace faithfully the
real mass averaged gas temperature. In Fig.~\ref{figure:tkin} we show
the average $N({\rm p}\ammo)$ and $T_{\rm kin}$ as functions of
distance from the core centre. The diagrams are derived by averaging
spectra over concentric, $10\arcsec$ wide annuli, and calculating the
parameters and their errors from these averaged spectra. Also shown in
this figure are the corresponding distributions of $N(\htwo)$ and
$T_{\rm C}$ derived from the Herschel/SPIRE maps. Two things are
perhaps worthy of notice in the diagrams: the $\ammo$ abundance seems
to decrease and the $T_{\rm kin}$ seems to increase towards the edge
of the core, although both quantities have large errors away
from the centre of the core. 

The line widths of the spectra observed towards the core centre range
from 170 to 260 m\,s$^{-1}$.  Assuming that the average kinetic
temperature in the core is $\sim 11$ K, the non-thermal velocity
dispersions obtained from various transitions are between $\sim 50$
and $\sim 100$ m\,s$^{-1}$.

\section{Abundances from radiative transfer modelling}
\label{sec:model}

\subsection{Physical model of H-MM1}
\label{ss:coremodel}

In order to account for inhomogeneities in the density and temperature
distributions along the line of sight, we constructed a spherically
symmetric physical model of the core. Besides providing a realistic
description of molecular line excitation conditions for the radiative
transfer modelling, the core model is also needed for making a
connection between the observations and the theory of interstellar
chemistry.  In this model the core structure is described by a
modified Bonnor-Ebert sphere (MBES) (\citealt{2001ApJ...557..193E};
\citealt{2001A&A...376..650Z}; \citealt{2011A&A...535A..49S};
\citealt{2015A&A...582A..48S}) which is a pressure bound, hydrostatic
sphere of gas and dust with the temperature decreasing towards the
centre.

We fixed the outer radius of the core to be $R_{\rm out} = 9600$ AU
($80\arcsec$). At this distance the core was assumed to be merged with
the ambient cloud. We assumed a constant non-thermal velocity
dispersion of $\sigma_{\rm NT}=100$ m\,s$^{-1}$ inside the core, which
increases the internal pressure slightly. The dust temperature profile
was calculated using a Monte Carlo program for continuum radiative
transfer,
CRT\footnote{\url{wiki.helsinki.fi/display/$\sim$mjuvela@helsinki.fi/CRT}},
developed by M. Juvela \citep{2005A&A...440..531J}. The spectrum of
the unattenuated interstellar radiation field (ISRF) was taken from
\cite{1994ASPC...58..355B}. We used the dust opacity data from
\cite{1994A&A...291..943O} for unprocessed dust grains with thin ice
coatings\footnote{\url{hera.ph1.uni-koeln.de/$\sim$ossk/Jena/tables.html}},
which agree with the opacities at 250, 350, and 500 $\mu$m used in the
derivation of the $T_{\rm C}$ and $N(\htwo)$ maps in Sect. 2.

The MBES model was constructed using the following constraints: 1) The
dust temperature at the boundary should agree with the $T_{\rm C}$
outside the core derived from Herschel; 2) the model should
approximately reproduce the 450 and 850 $\mu$m emission profiles
derived from SCUBA-2 maps of \cite{2015MNRAS.450.1094P}; 3) the
mass-averaged gas kinetic temperature profile, smoothed to the angular
resolution of the GBT, should agree with the observed $T_{\rm kin}$
profile shown in Fig.~\ref{figure:tkin}. In order to achieve the large
temperature difference between the edge ($\sim 15$ K) and the centre
($<11$ K) with the model where the external heating is dominated by
the ISRF, we set the visual extinction to $A_{\rm V} = 2^{\rm mag}$ at
the outer boundary of the core. With this choice we assume that
  the core lies near the edge or in a protrusion of the ambient cloud. The total
  hydrogen column density in the neighbourhood of the core is of the
  order of $10^{22}$ cm$^{-2}$ (Fig.~\ref{figure:hmm1_td_nh2_maps})
  which corresponds to $A_{\rm V} \sim 10^{\rm mag}$.  The strongly
peaked sub-millimetre emission observed with SCUBA-2 could only be
reproduced with central $\htwo$ densities of the order of $10^6$
cm$^{-3}$.

The iteration was started from the density distribution of an
isothermal Bonnor-Ebert sphere at 11 K with a central density of
$n(\htwo)=10^6$ cm$^{-3}$. The intensity of the ISRF and the central density
were adjusted to reach agreement with the constraints 1) and 2) above.
The sub-millimetre intensity profiles of the model core were
calculated by evaluating the integrals $\int_0^L B_\nu(T_{\rm dust},z)
\rho(z) \kappa_\nu dz$, where the integration is along the line of
sight, at different distances from the core centre. The resulting
intensity profiles at $450\,\mu$m and $850\,\mu$m, smoothed to
appropriate resolutions, were compared with corresponding circularly
averaged profiles from the SCUBA-2 maps.

The gas and dust temperatures are assumed to be equal at densities
above $\sim 10^5$ cm$^{-3}$ \citep{2001ApJ...557..736G}. The observed
$T_{\rm kin}$ profile shown in Fig.~\ref{figure:tkin} suggests,
however, that the (3-dimensional) gas temperature distribution inside
the core does not show the steep gradient characteristic of $T_{\rm
  dust}$ caused by the attenuation of the ISRF. In fact, the ammonia
observations can be explained with a model where the core is mostly
isothermal, but the temperature rises steeply near the edge. We note
that this conclusion is probably influenced by the limited angular
resolution (cf.  temperature determination in L1544 by
\citealt{2007A&A...470..221C}).  To conform with the prediction that
$T_{\rm gas} \sim T_{\rm dust}$ at the highest densities, we assumed
that this is true above a certain, adjustable density threshold, but
that below this threshold, $T_{\rm gas}$ is constant up to the
transition layer, where it rises abruptly.  The adopted gas
temperature distribution required a slight adjustment of the density
profile to keep the core in hydrostatic equilibrium. The iteration
converged after three-four rounds.  A reasonable agreement with the
dust continuum observations was found with a model where the central
density of core is $n(\htwo)=1.2\,10^6$ cm$^{-3}$, and the standard
IRSF is scaled up by the factor 1.7. The observed $T_{\rm kin}$
distribution could be reproduced assuming that $T_{\rm gas}$ separates
from $T_{\rm dust}$ at the density $n(\htwo) \sim 4\,10^5$ cm$^{-3}$.
At this point, $\sim 15\arcsec$ from the centre, the temperature is
$\sim 11$ K.

\begin{figure}
\includegraphics[width=9cm]{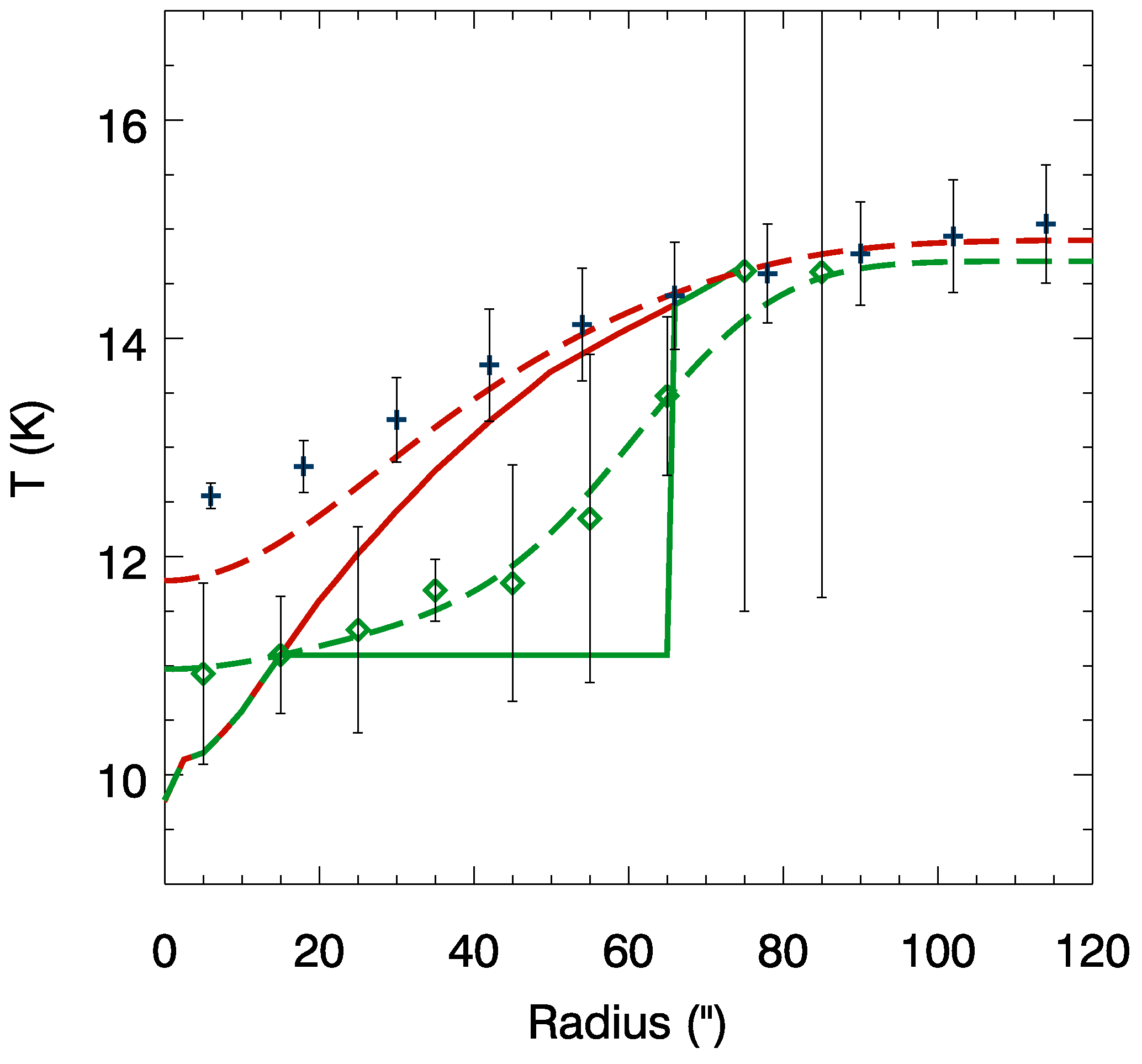}

\caption{Radial $T_{\rm dust}$ (red solid curve) and
  $T_{\rm gas}$ (green solid curve) distributions of the hydrostatic
  core model of H-MM1 used in the chemistry modelling and radiative
  transfer calculations. The corresponding mass-averaged dust and gas
  temperature profiles are shown as
  dashed red and green curves. The $T_{\rm C}$ distribution derived
  from Herschel data is indicated with black plus signs, and the
  $T_{\rm kin}$ distribution from the GBT ammonia data is indicated
  with green diamonds.}
\label{figure:be_tdust}
\end{figure}

\begin{figure}
\includegraphics[width=9cm]{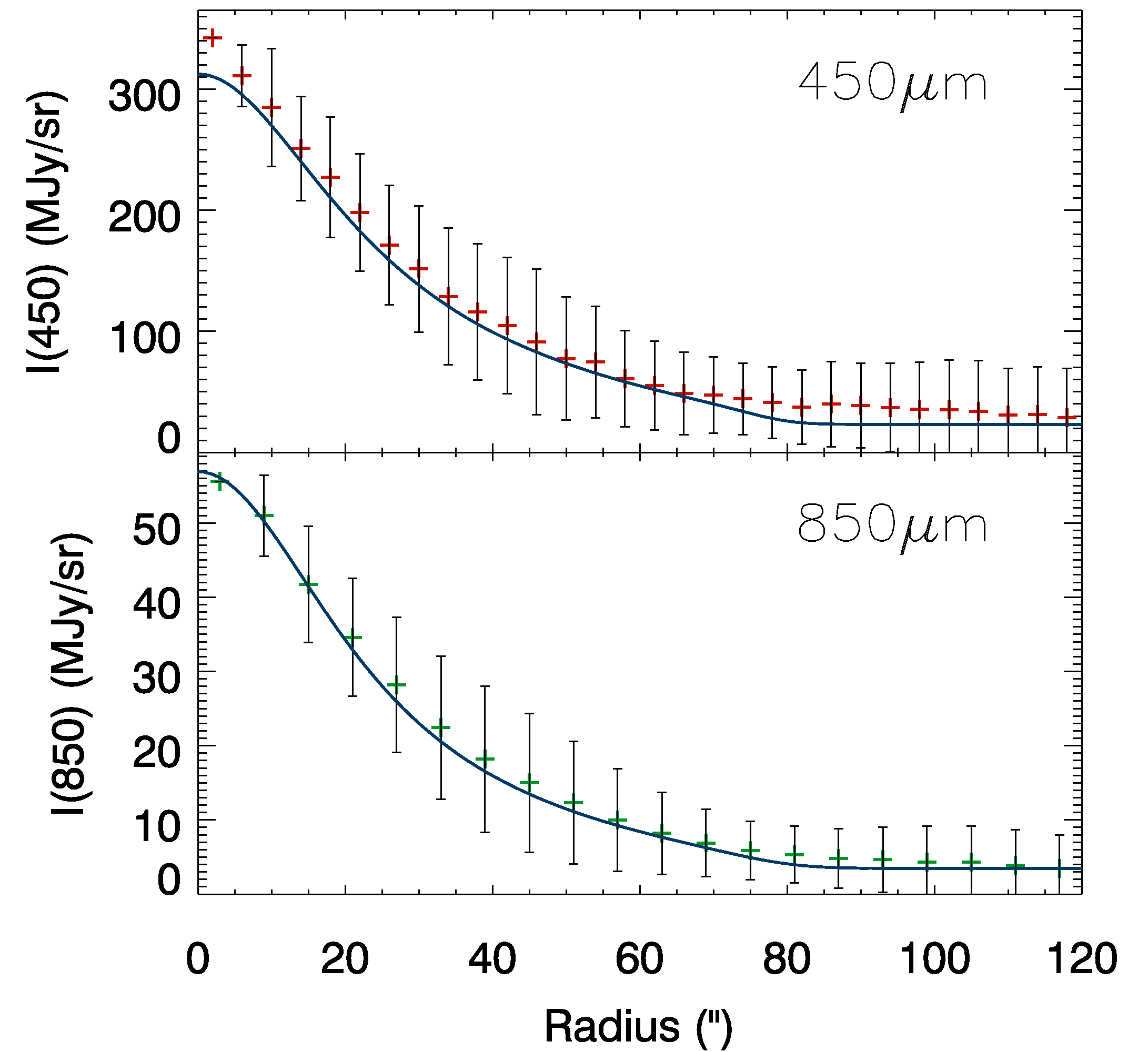}
\caption{Sub-millimetre intensities as functions of radial distance
  from the centre of H-MM1. The plus signs with error bars indicate
  averages over concentric annuli and their standard deviations. These
  are obtained from SCUBA-2 maps at $450\,\mu$m and $850\,\mu$m
  published by \cite{2015MNRAS.450.1094P}. The solid curves are
  predictions from the MBES model described in the text and in
  Fig.~\ref{figure:be_tdust}.}
\label{figure:iprofs}
\end{figure}

The distributions of $T_{\rm gas}$ and $T_{\rm dust}$ as functions of
the radial distance from the core centre for the best fit model are
shown in Fig.~\ref{figure:be_tdust}, together with the observable
mass-averaged temperature profiles smoothed to the angular resolutions
of GBT and Herschel.  Fig.~\ref{figure:iprofs} shows the circularly
averaged $450\,\mu$m and $850\,\mu$m intensity profiles of H-MM1
derived from the SCUBA-2 maps of \cite{2015MNRAS.450.1094P}, together
with predictions from the MBES model.

\subsection{Average abundances from line modelling}
\label{ss:ave_abus}

We derived the fractional abundances of the observed molecules in
H-MM1 by applying the Monte Carlo radiative transfer program of
\cite{1997A&A...322..943J} to the physical model described in 
Sect.~\ref{ss:coremodel}.  The collisional rate coefficients for
$\dammo$ were adopted from \cite{2014MNRAS.444.2544D}, and for
$\ddammo$ and $\ndthree$ we used the newly calculated coefficients
from \cite{2016MNRAS.457.1535D}.  For $\ddiaz$ we have used the
collisional rate coefficients for $\diaz$ from the recent work of
\cite{2015MNRAS.446.1245L}.

In this calculation, we assumed constant fractional abundances
throughout the core, i.e., that there is no dependence on the distance
from the core centre. The obtained values can thus be taken as
averages over the core. Starting from the estimates presented in
  Table~\ref{table:hffits} of Sect.~\ref{sec:LTE_analysis}, we varied
  fractional abundances until the modelled spectra produced the same
  integrated intensities as the observed ones.  The best-fit
  fractional abundances are presented in Table~\ref{table:const_abus}.
  The uncertainties correspond to the 1-$\sigma$ errors of the
  integrated intensities. The {\sl para}-$\ammo$ lines are broader
  than those of its deuterated isotopologues, and for this molecule we
  had to increase the assumed non-thermal velocity dispersion to
  $\sigma_{\rm NT}=150$ m\,s$^{-1}$ in order to reproduce the
  integrated intensities. This suggests that $\ammo$ emission has
  contribution from the ambient cloud having a larger velocity
  dispersion than the core, and that the derived fractional 
  {\sl para}-$\ammo$ abundance is an upper limit for the core.

For the different isotopologues of ammonia, the agreement reached
between the predicted spectra and observations is equally good as for
the hyperfine fits shown in Figs.~\ref{figure:ammo} and
\ref{figure:dammo}. The two $\ddiaz$ lines detected cannot be
reproduced by a single abundance. The value listed in
Table~\ref{table:const_abus} is a compromise which overpredicts the
$\ddiaz(4-3)$ intensity but gives too a weak $\ddiaz(2-1)$ line. The
situation is thus opposite to what is expected from the results of the
hyperfine fits. This indicates that either the assumption of a
constant abundance is unrealistic for $\ddiaz$ or that the physical
model is inaccurate.

\begin{table}
  \caption{Fractional abundances, deuterium fractionation ratios, and 
    spin ratios in H-MM1 derived from the detected lines using radiative 
transfer modelling. The abundances are assumed to be constant through the core.}
\begin{center}
\begin{tabular}{ll}  \hline\hline
p$\ammo$ & $(3.8 \pm 0.1)\,10^{-9}$  \\
o$\dammo$ & $(2.2 \pm 0.1)\,10^{-9}$  \\
p$\dammo$ & $(7.3 \pm 0.3)\,10^{-10}$ \\
o$\ddammo$ & $(4.5\pm 0.3)\,10^{-10}$ \\
p$\ddammo$ & $(1.9\pm 0.2)\,10^{-10}$ \\
m$\ndthree$ & $(1.4\pm 0.1)\,10^{-11}$ \\
$\ddiaz$  & $(6.6\pm2.3)\,10^{-11}$ \\ \hline
 $\dammo/\ammo$ & $0.39 \pm 0.02$ (o:p$\ammo$  = 1:1) \\
 $\ddammo/\dammo$ &  $0.22\pm0.02$ \\ 
$\ndthree/\ddammo$ & $0.06 \pm 0.01$ (o:m:p$\ndthree$ = 18:10:1) \\ \hline  
o/p-$\dammo$ & $3.0\pm 0.2$ \\
o/p-$\ddammo$ & $2.4\pm 0.4$ 
\end{tabular}
\end{center}
\label{table:const_abus}
\end{table}

The abundances listed in Table~\ref{table:const_abus} imply the
following total fractional abundances: $X(\ammo)= (7.6\pm
0.1)\,10^{-9}$ (assuming o:p=1:1), $X(\dammo) = (2.9 \pm 0.1)
\,10^{-9}$, $X(\ddammo) = (6.4 \pm 0.5)\,10^{-10}$, and $X(\ndthree)=
(3.7\pm 0.4)\,10^{-11}$ (assuming o:m:p = 18:10:1). Here we have
assumed statistical spin ratios for species without a line
detection. The corresponding deuterium fractionation ratios for
ammonia, and the ortho/para ratios for $\dammo$ and $\ddammo$ are
listed in the bottom part of Table~\ref{table:const_abus}.

For the molecules with the brightest lines, the abundances from
  the radiative transfer modelling agree relatively well with those
  from the LTE analysis (Table~\ref{table:hffits}). The most glaring
  discrepancy (by a factor of 2.5) is found for {\sl para}-$\ddammo$
  with the weakest detection and a very large uncertainty of the
  optical thickness from the LTE method. The statistical errors of the
  fractional abundances listed in Table~\ref{table:const_abus} are
  small in most cases, but the values are subject to systematic
  errors depending on accuracy of the physical model, and on the
  validity of the assumption of constant abundances.  The present
  physical model is, however, consistent with dust continuum
  observations, and provides a more realistic decription of the
  excitation conditions in the core than the assumption of
  line-of-sight homogeneity. Therefore, we consider the abundance
  ratios listed in Table~\ref{table:const_abus} to be more accurate
  than those implied by the values presented in
  Table~\ref{table:hffits}, and use the former ratios to assess the
  validity of the chemistry model described below.

\section{Chemical modelling}

\subsection{Model description}
\label{ss:chemistry_model}

We model the chemistry of H-MM1 using the pseudo-time-dependent
gas-grain chemical code presented in earlier papers \citep{Sipila12,
  2013A&A...554A..92S, 2015A&A...578A..55S, 2015A&A...581A.122S} where
the details of the code (e.g., the expressions of the various reaction
rate coefficients) can be found. The model includes gas-phase
chemistry, adsorption onto and (non-thermal) desorption from grain
surfaces, and grain-surface chemistry. Tunnelling diffusion of H and D
atoms on grains is not considered in the present calculations, whereas
tunnelling though activation energy barriers in surface reactions has
been included. 

The program can be instructed to include various desorption
mechanisms: thermal desorption (negligible in the physical conditions
explored here), cosmic ray desorption, reactive desorption, and
photodesorption. Cosmic ray desorption is treated following
\cite{HH93}. Exothermic association reactions on the surface can
result, when this option is turned on, in desorption of the reaction
product with an efficiency of 1\% \citep{Garrod07}.  Finally, it is
possible to include photodesorption of water, CO, and ammonia caused
by secondary UV photons created by $\rm H_2$ excitation
\citep{Prasad83}. In the simulations presented here, however, the
photodesorption and reactive desorption options have been turned
off. According to extensive testing, the inclusion of these processes
increases the ammonia production, which is turn makes it necessary to
lower the elemental N abundance to ensure compliance with the observed
line intensities, but fractionation and spin ratios remain largely
unchanged.

The spin-state chemical model presented in \citet{2015A&A...578A..55S}
describes the spin states of species involving multiple
protons. Recently, we upgraded this model to include a self-consistent
description of the spin states of multiply-deuterated species
\citep{2015A&A...581A.122S}. This is achieved by considering nuclear
spin selection rules arising from molecular symmetries, assuming full
scrambling of nuclei in reactive collisions. The model gives the
necessary information for the present application, i.e., the
calculation of simulated line emission from deuterated ammonia.  In
the beginning of the simulation, all elements are in the atomic form,
with the exceptions of hydrogen and deuterium which are initially
locked in $\htwo$ and HD, respectively. 

The descriptions of deuterium and spin-state chemistry adopted here
apply to both gas-phase chemistry and grain-surface
chemistry. However, the formation mechanism of ammonia (as a
particular example) is different in the gas and on the grains. In the
gas phase, (deuterated) ammonia forms through a network of
ion-molecule reactions, while on the surface the formation mechanism
is hydrogen/deuterium addition. Therefore we expect non-statistical
deuterium and spin-state ratios in the gas phase, and statistical
ratios on the grain surface, although complete scrambling is assumed
in both cases. The assumed binding energies on grain surfaces are the
same as listed in Table~2 of \cite{2015A&A...578A..55S}.

For the elemental abundances we have adopted the set of low-metal
elemental abundances labelled EA1 in \cite{2008ApJ...680..371W},
except for the nitrogen abundance for which we needed to increase
the EA1 abundance by a factor of 2.5 to N/H=$5.3\,10^{-5}$ to reproduce
the observed line intensities of $\dammo$, $\ddammo$, $\ndthree$, and 
$\ddiaz$. According to the recent compilation of \cite{2009ApJ...700.1299J},
the appropriate value for the diffuse ISM is N/H$=6.2\,10^{-5}$. 
The adopted carbon and oxygen abundances are O/H$=1.8\,10^{-4}$, 
C/H$=7.3\,10^{-5}$.

We derive abundance profiles for the various molecules by separating
the core model (see Sect.\,\ref{ss:coremodel}) into a series of
concentric spherical shells, each associated with unique values of
density, gas/dust temperatures, and visual extinction $A_{\rm V}$.
Chemical evolution is then calculated separately in each shell,
which leads to simulated abundances for each chemical species as
functions of time and radial distance from the core centre. The
integration of the modelled abundance gradients into the radiative
transfer model is discussed in Sect.\,\ref{ss:RT}.

\subsection{Chemical evolution of the core}

\label{ss:chem_evolution}

We calculate the evolution of chemical abundances in the core model
with the goal to examine if the model can reproduce, at a certain
stage of the simulation, the intensities of the lines of deuterated 
ammonia and $\ddiaz$ observed towards H-MM1 in the present
study, as well as the intensities of the o$\htwod$ and p$\dtwoh$ lines
observed previously by \cite{2011A&A...528C...2P}.

We assume that the spin temperature of $\htwo$ has been thermalized
during the intial contraction phase of the cloud down to $\sim 20$ K
(\citealt{2006A&A...449..621F}; \citealt{2013A&A...554A..92S}), and
accordingly, set the initial o/p$\htwo$ to $1\,10^{-3}$.  For the
cosmic ray ionisation rate of $\htwo$ we assume
$\zeta_{\htwo}=1.3\,10^{-17}$ s$^{-1}$, and the average grain radius
is set to $a=0.1\,\mu$m. These are kept unchanged in the present
simulations. We note, however, that deuteration can be delayed by
increasing the initial o/p$\htwo$ ratio, and that the fractionation
ratios can be lowered by decreasing the average grain size or by
increasing the cosmic ionization rate \citep{2010A&A...509A..98S}. On
the other other hand, an increase of the cosmic ray ionisation rate
would generally increase the abundances and the line intensities of
$\hthree$, $\ammo$, and their deuterated isotopologues, and decrease
the abundances and line intensities of $\diaz$ and $\ddiaz$.  

\begin{figure}
\includegraphics[width=9cm]{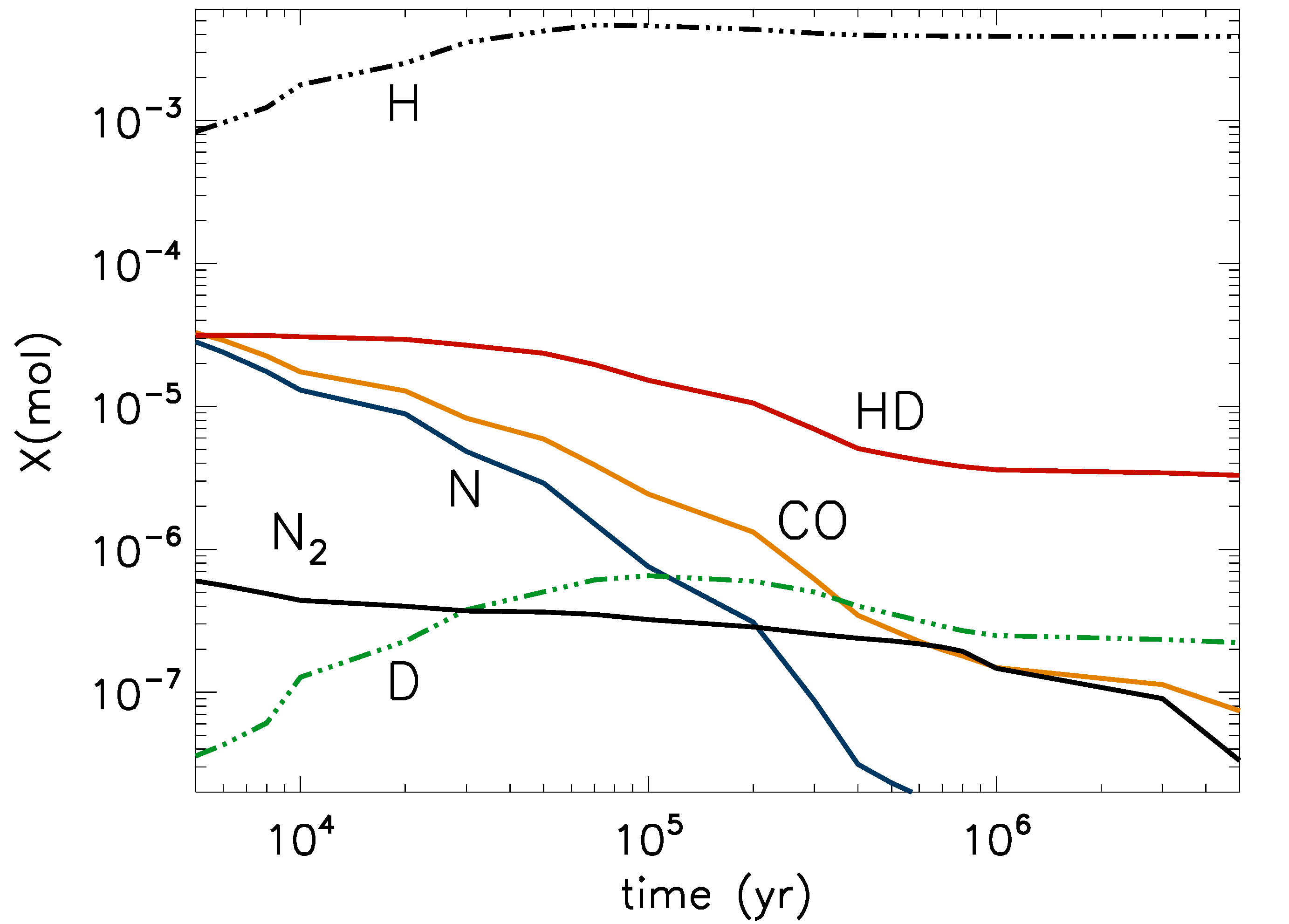}
\caption{Fractional abundances of selected species relative to $\htwo$ 
 as functions of time in the core model described 
 in Sect.~\ref{ss:coremodel}. 
The abundances are density weighted averages.}
\label{figure:co_etc}
\end{figure}

We first discuss the predictions for some of the most common
species. The gas-phase abundances of the H, D, and N atoms, and the
HD, CO, and ${\rm N_2}$ molecules, relative to the total hydrogen
abundance, are plotted in Fig.~\ref{figure:co_etc}, as functions of
time for our fiducial core model discussed in
Sect.~\ref{ss:coremodel}. The abundances are averages over the line of
sight through the centre of the core, weighted by the density.  The
freeze-out of CO is followed by an increase of $\hthree$, which in
turn results in an enhanced {\sl ortho}-{\sl para} conversion of
$\htwo$. At the same time, deuterium is efficiently transferred from
HD to deuterated ions in the gas phase (to $\htwod$, $\dtwoh$, and
$\dthree$ in the first place).  The H and D atoms released in the
dissociative recombination of deuterated ions mainly accrete onto
grains, where they can combine to give back $\htwo$ or HD, but also
react with heavier atoms or radicals. At late stages of chemical
evolution, deuterium becomes increasingly incorporated in icy
compounds. In the gas phase this is reflected by the reduction of the
HD abundance.

The rapid decrease of atomic nitrogen in the beginning of the
simulation is mainly caused by accretion onto grains. A fraction of
the nitrogen atoms in the gas phase is converted to N$_2$ through
${\rm N} + {\rm OH} \rightarrow {\rm NO} + {\rm H}$, ${\rm N} + {\rm
  NO} \rightarrow {\rm N_2} + {\rm O}$ (\citealt{2006A&A...456..215F};
\citealt{2014A&A...562A..83L}). This reaction is important at early
times, when the N and OH abundances are high. Also the ${\rm N_2}$
molecules accrete onto grains, but their desorption from grains is
more significant than for N atoms, which quickly react with other
atoms or radicals, e.g., ${\rm N^*} + {\rm H^*} \rightarrow {\rm
  NH^*}$, ${\rm N^*} + {\rm O^*} \rightarrow {\rm NO^*}$, or ${\rm
  N^*} + {\rm N^*} \rightarrow {\rm N_2^*}$. Species attached to grain
are indicated here with asterisks.  The surface species ${\rm N_2^*}$
is destroyed by two processes only, either by desorption or by
photodissociation, ${\rm N_2^*} + \, \mbox{photon} \, \rightarrow {\rm
  N^*} + {\rm N^*}$, by cosmic-ray induced UV photons. This is the
main difference from ${\rm CO^*}$ for which hydrogenation, ${\rm CO^*}
+ {\rm H^*} \rightarrow {\rm HCO^*}$, competes hard against desorption
(see also Sect.~\ref{ss:ubiquity}).

Owing to desorption, the ${\rm N_2}$ abundance remains high in the gas
phase until late stages of the simulation.  Molecular nitrogen is
prerequisite to $\diaz$ which forms through ${\rm N_2} + \hthree
\rightarrow \diaz + \htwo$. In case reactive desorption is included,
the most important source of ammonia at very early stages of 
simulation is formation on grains, ${\rm NH_2^*} + {\rm H^*}
\rightarrow \ammo$, where ammonia is supposed to be released into the
gas phase at the probability of $1\%$.  After a few thousand years,
gas-phase formation takes over. In the simulations presented here, the
ammonia production is always dominated by the well-known chain of
gas-phase reactions, terminating in $\ammohplus + \el \rightarrow
\ammo + {\rm H}$ (e.g., \citealt{2014A&A...562A..83L}; 
\citealt{2015A&A...576A..99R}). As discussed
by \cite{2015A&A...581A.122S}, the dissociative ionisation of ${\rm
  HNC}$ by ${\rm He^+}$ helps the initiation of this chain by
producing ${\rm NH^+}$, which otherwise would be solely dependent on
${\rm N^+} + {\rm o}\htwo \rightarrow {\rm NH^+} + {\rm H}$
\citep{2012A&A...537A..20D}.

\vspace{2mm}

\begin{figure}
\includegraphics[width=9.3cm]{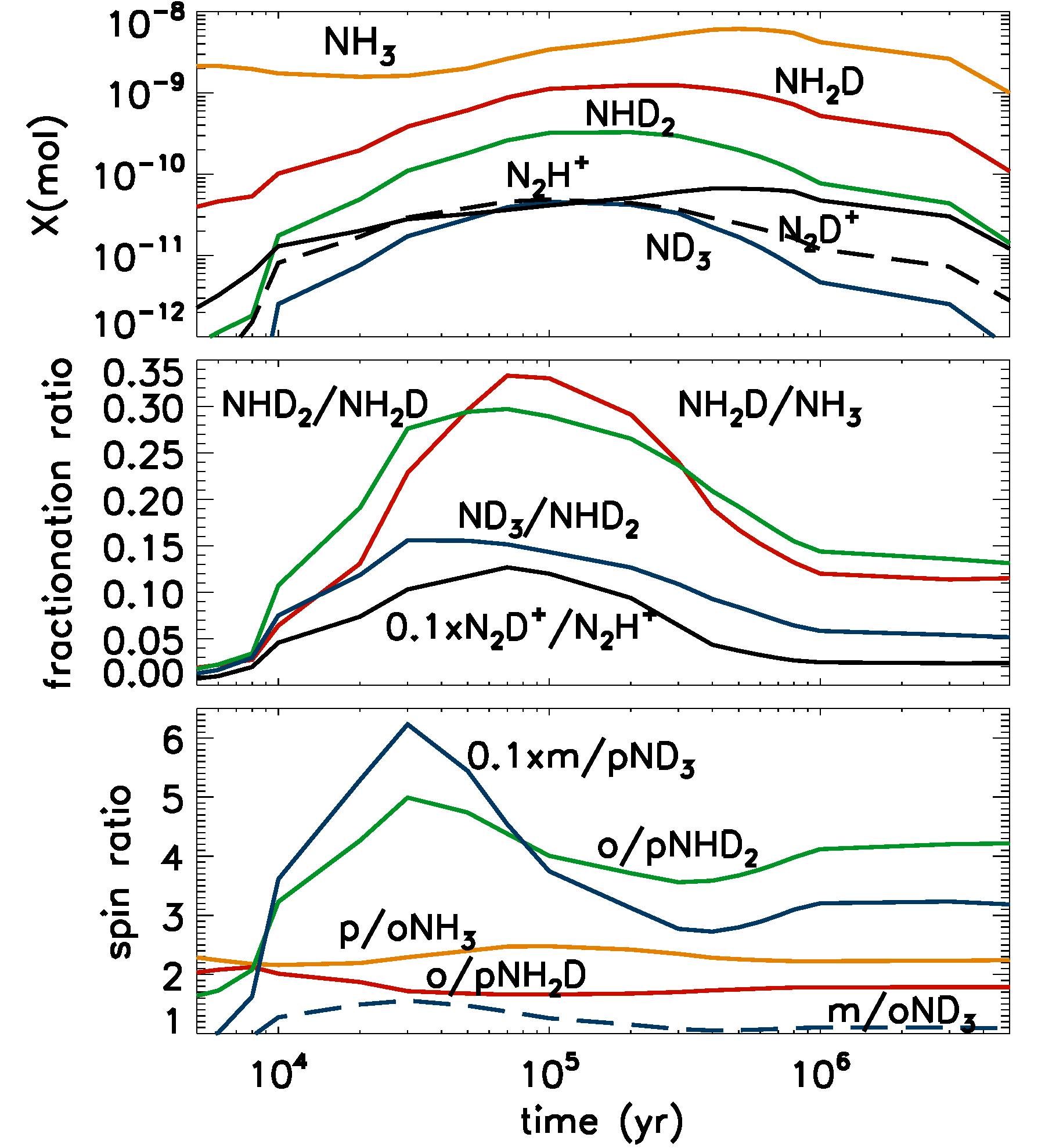}
\caption{Gas-phase fractional abundances of $\ammo$ and $\diaz$ and 
 their deuterated isotopologues as functions of time in
  the core model. The
   $\ddiaz/\diaz$ and m/p$\ndthree$ ratios are divided by 10 to make
   the other ratios readable in these diagrams.}
\label{figure:abu_a_vs_time}
\end{figure}

\begin{figure}
\includegraphics[width=9.3cm]{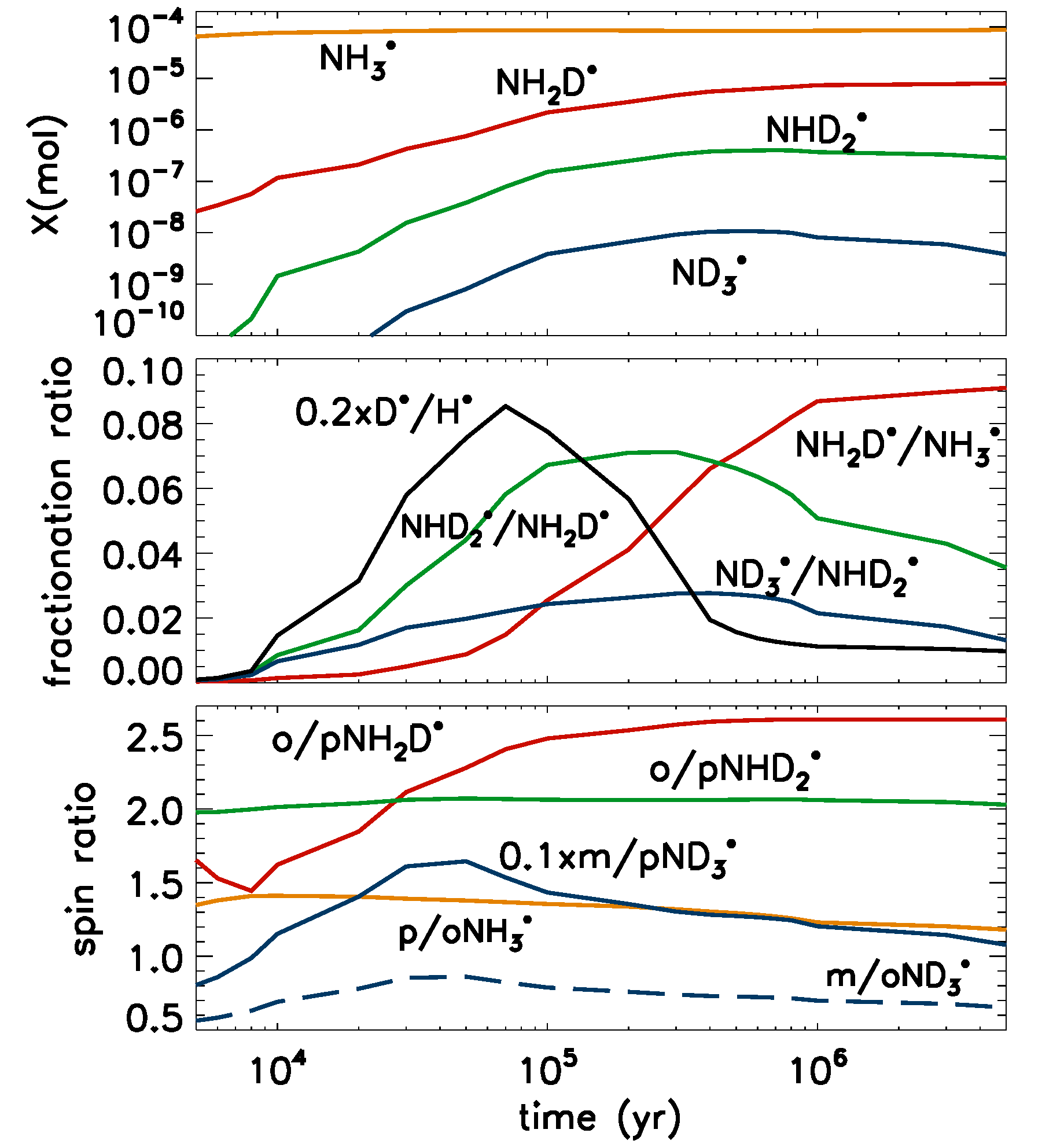}
\caption{Evolution of the abundances of $\ammo$ and 
 its deuterated isotopologues on grain surfaces for 
  the core model. The
   m/p$\ndthree$ ratios are divided by 10. The 
  D/H ratio on grains shown in the middle panel is multiplied
  by 0.2.}
\label{figure:abu_c_vs_time}
\end{figure}

The evolution of the abundances of $\ammo$ and $\diaz$, and their
deuterated isotopologues in the gas phase are shown in
Fig.~\ref{figure:abu_a_vs_time} (top panel). The other two panels show
the fractionation ratios and the spin ratios for these species. In
Fig.~\ref{figure:abu_c_vs_time}, we show the abundances, fractionation
ratios, and the spin ratios of the grain-surface species $\ammo^*$,
$\dammo^*$, $\ddammo^*$, and $\ndthree^*$. The middle panel of this
latter figure also shows the atomic D$^*$/H$^*$ on grains.

\begin{figure}
\includegraphics[width=9.3cm]{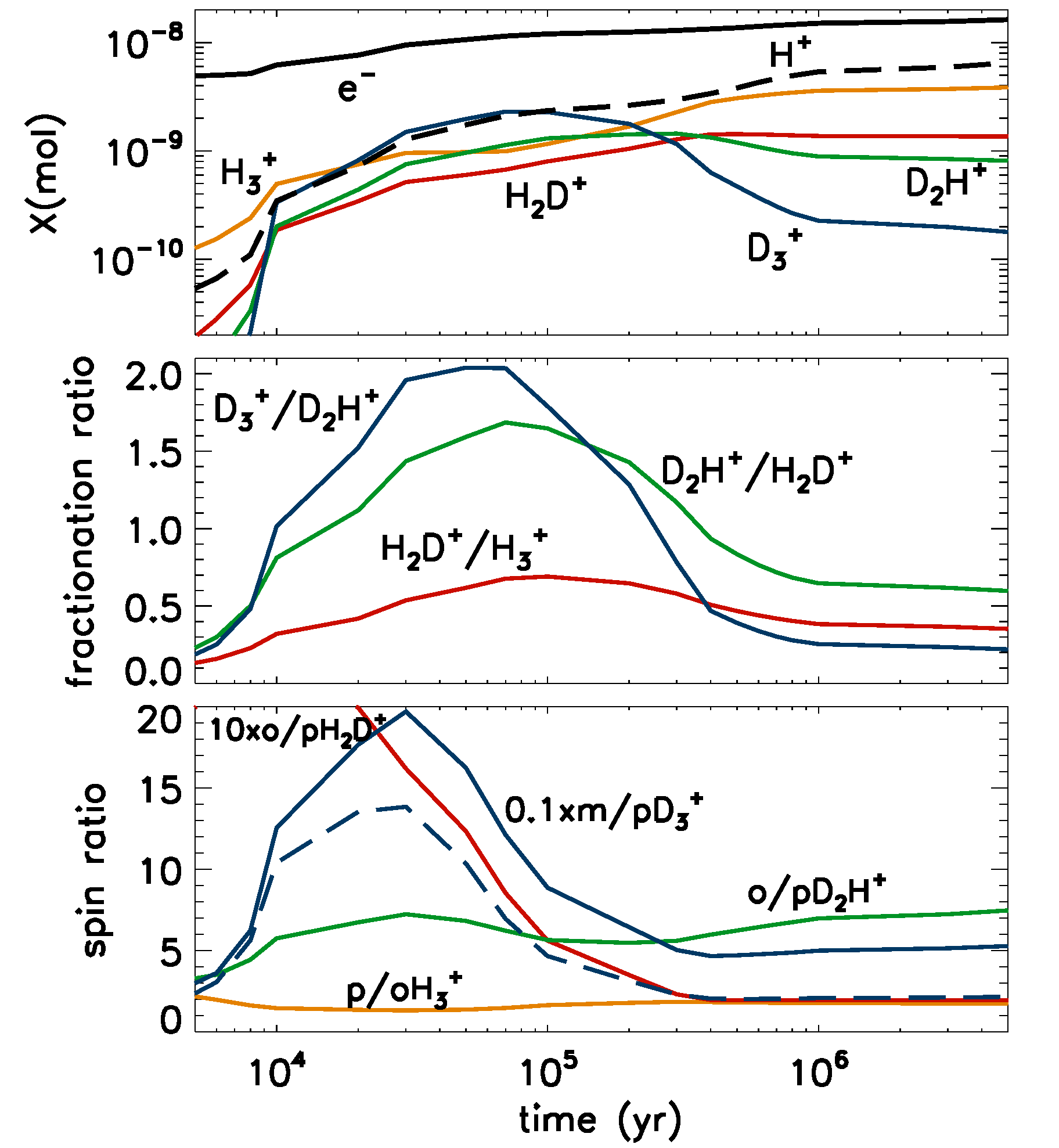}
\caption{Evolution of the isotopologues of $\hthree$ in the core model.  
  The m/p$\ndthree$ ratio is divided by 10.}
\label{figure:abu_b_vs_time}
\end{figure}

In the present model, the gas-phase $\ammo$ abundance is built up
early, and it does not change significantly at later times.  The
largest variations are seen in the abundances of $\ddammo$,
$\ndthree$, and $\ddiaz$, which grow rapidly in the beginning, and
decay slowly after the deuteration peak. This behaviour seems to
reflect the variations in the $\dtwoh$ and $\dthree$ abundances which
are shown in Fig.~\ref{figure:abu_b_vs_time}.  The $\dthree$ ion
reaches its maximum before $\dtwoh$, which in turn peaks before
$\htwod$. Likewise, the maximum fractionation ratio $\ndthree/\ddammo$
occurs earlier than the maximum in the $\ddammo/\dammo$ ratio, which
again takes place long before the $\dammo/\ammo$ peak. 

In the bottom panel of Fig.~\ref{figure:abu_a_vs_time} one can see
that while the o/p-$\dammo$ ratio decreases slightly with time,
o/p-$\ddammo$, m/p-$\ndthree$, and m/o-$\ndthree$ have increasing
tendencies.  The predicted ratios are close to their statistical
values in the beginning of the simulation. The o/p-$\ddammo$ and
m/p-$\ndthree$ ratios mimic the corresponding ratios of $\dtwoh$ and
$\dthree$ shown in Fig.~\ref{figure:abu_b_vs_time}, bottom panel.  The
relationship between the spin modifications of $\ammo$ and $\hthree$
is discussed in Sect.~6.

The ammonia production on grain surfaces is very efficient in the
present model. The deuteration of ammonia occurs more slowly than in the gas
phase, and never reaches as large fractionation ratios as seen there.
The spin ratios on grains stay close to their statistical values at
all times, except that m$\ndthree^*$ is enhanced at the cost of
o$\ndthree^*$ and p$\ndthree^*$. 

\subsection{Predicted spectra}
\label{ss:RT}

The radial distributions of the density, temperature, and the
(time-varying) chemical abundances in the gas phase are used as input
for a Monte Carlo radiative transfer program
\citep{1997A&A...322..943J} to predict observable rotational line
profiles. Like in simulations described in
  Sect.~\ref{ss:ave_abus}, we use a larger non-thermal velocity
  dispersion ($\sigma_{\rm N.T.} = 150\, {\rm ms}^{-1}$) for $\ammo$
  than for the deuterated species (for which $\sigma_{\rm N.T.} =
  100\, {\rm ms}^{-1}$) in order to reach agreement with the observed
  integrated intensities.
 
\begin{figure}
\includegraphics[width=9.3cm]{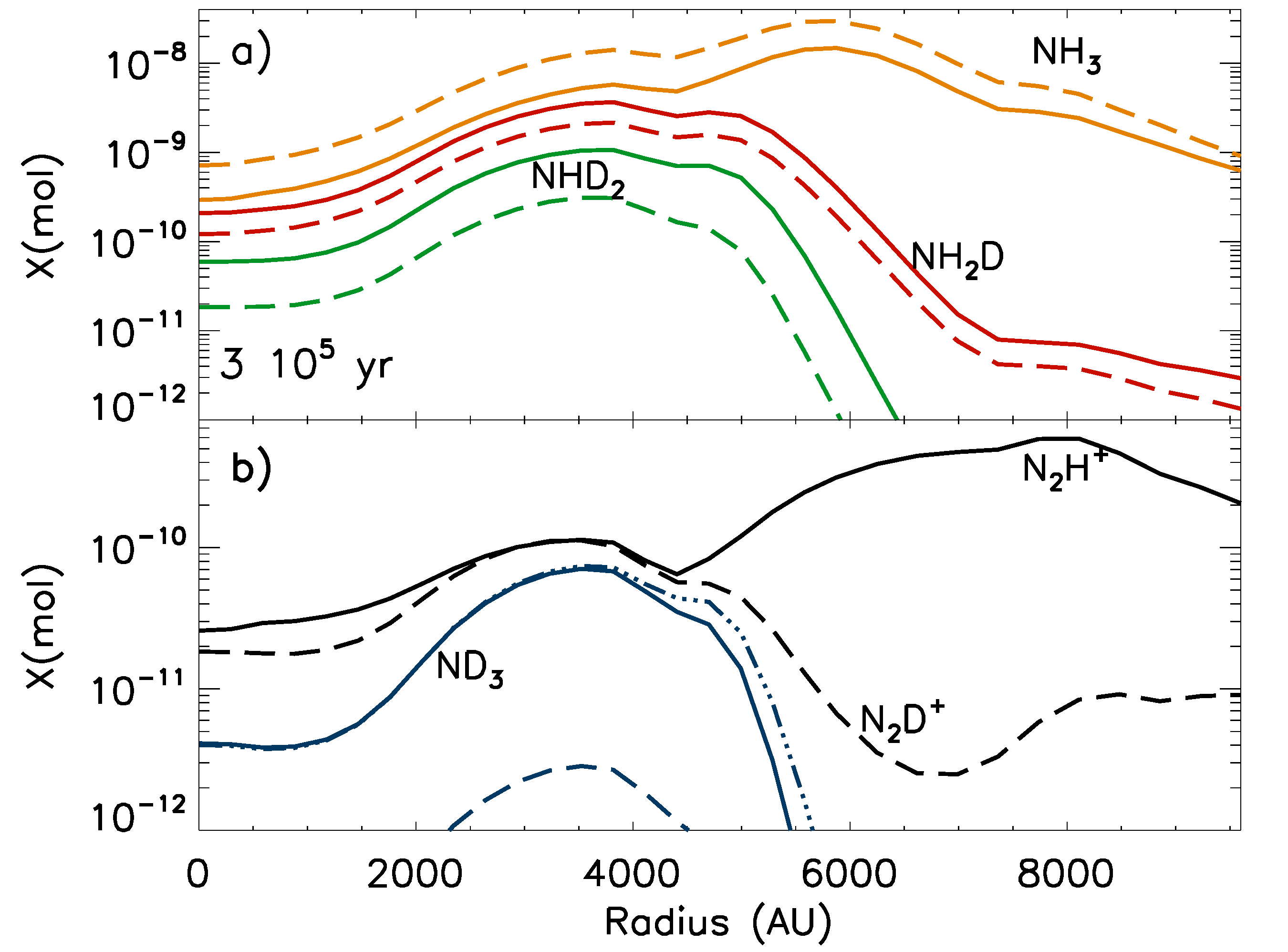}
\includegraphics[width=9.3cm]{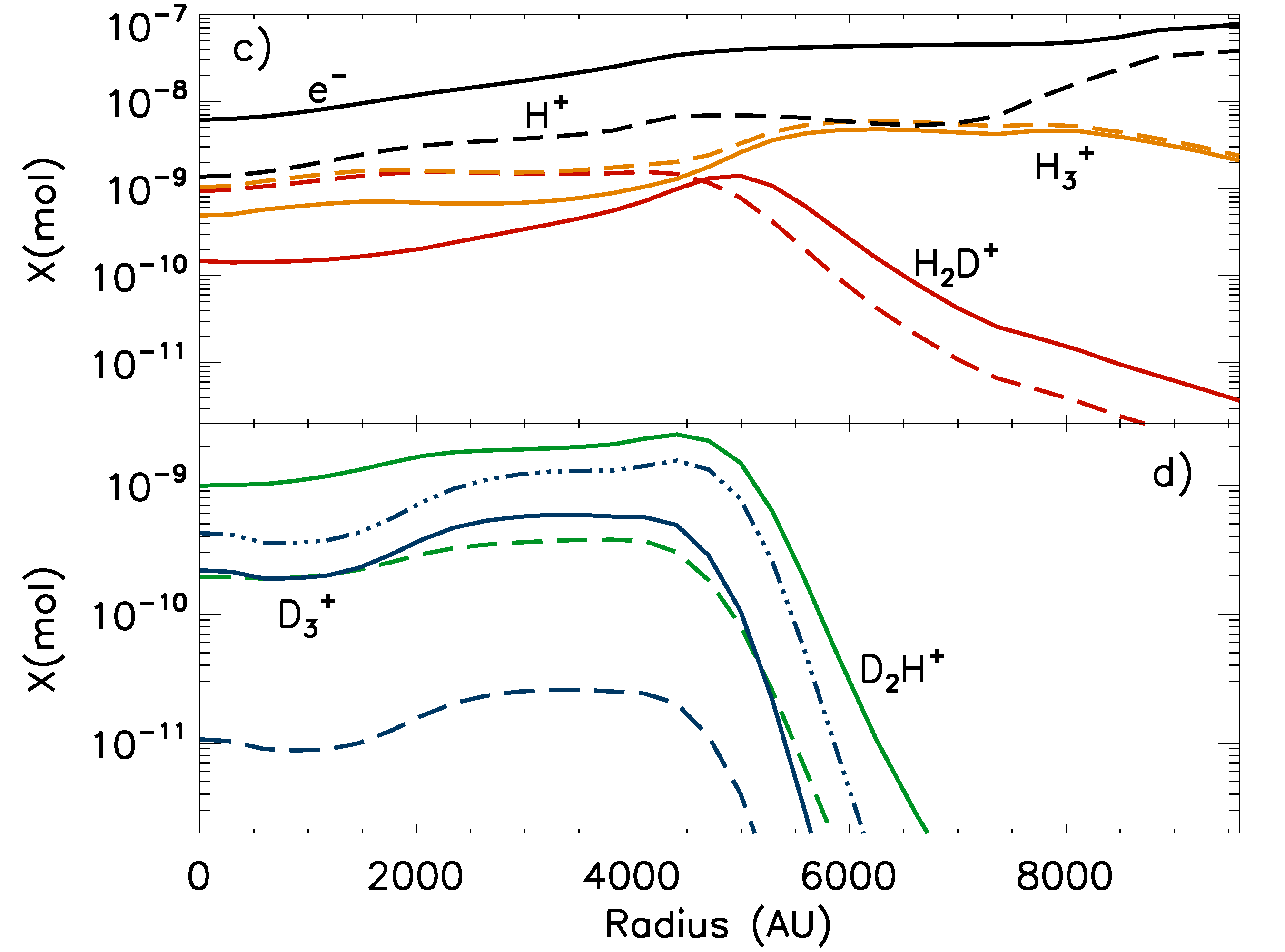}
\caption{Fractional abundances of selected species as
  functions of the radial distance from the core centre at the time
  $3\,10^5$ yr from the beginning of the simulation. The abundances of 
  {\sl ortho} species are drawn with solid lines, and those of  
  {\sl para} species are drawn with dashed lines. The abundances of 
   {\sl meta} species ($\ndthree$ and $\dthree$) are indicated with dash-dotted
   curves.}
\label{figure:abu_vs_rad}
\end{figure}

The line intensities depend, besides the (mass averaged)
abundances of the species within the telecope beam, also on their
radial distributions, which change with time. At early stages, 
deuterated species are concentrated on the core centre with high
densities, whereas later on, when freezing onto grains reduces the
abundances in the centre, line emission is dominated by lower-density
outer parts of the core. The changes of the abundance profiles cause
that at early times, lines with large transition dipole moments like
$\ddiaz(4-3)$ and m$\ndthree(1_0-0_0)$ are much stronger than the
o/p$\dammo(1_{11}-1_{01})$ lines, for example, but the reverse is true at
later times. The fractional abundances of selected species as
functions of the radius are shown in Fig.~\ref{figure:abu_vs_rad}. The
distributions are taken at the time when most of the predicted spectra
agree reasonably well with the observations.

\vspace{2mm}

\begin{figure}
\includegraphics[width=9.3cm]{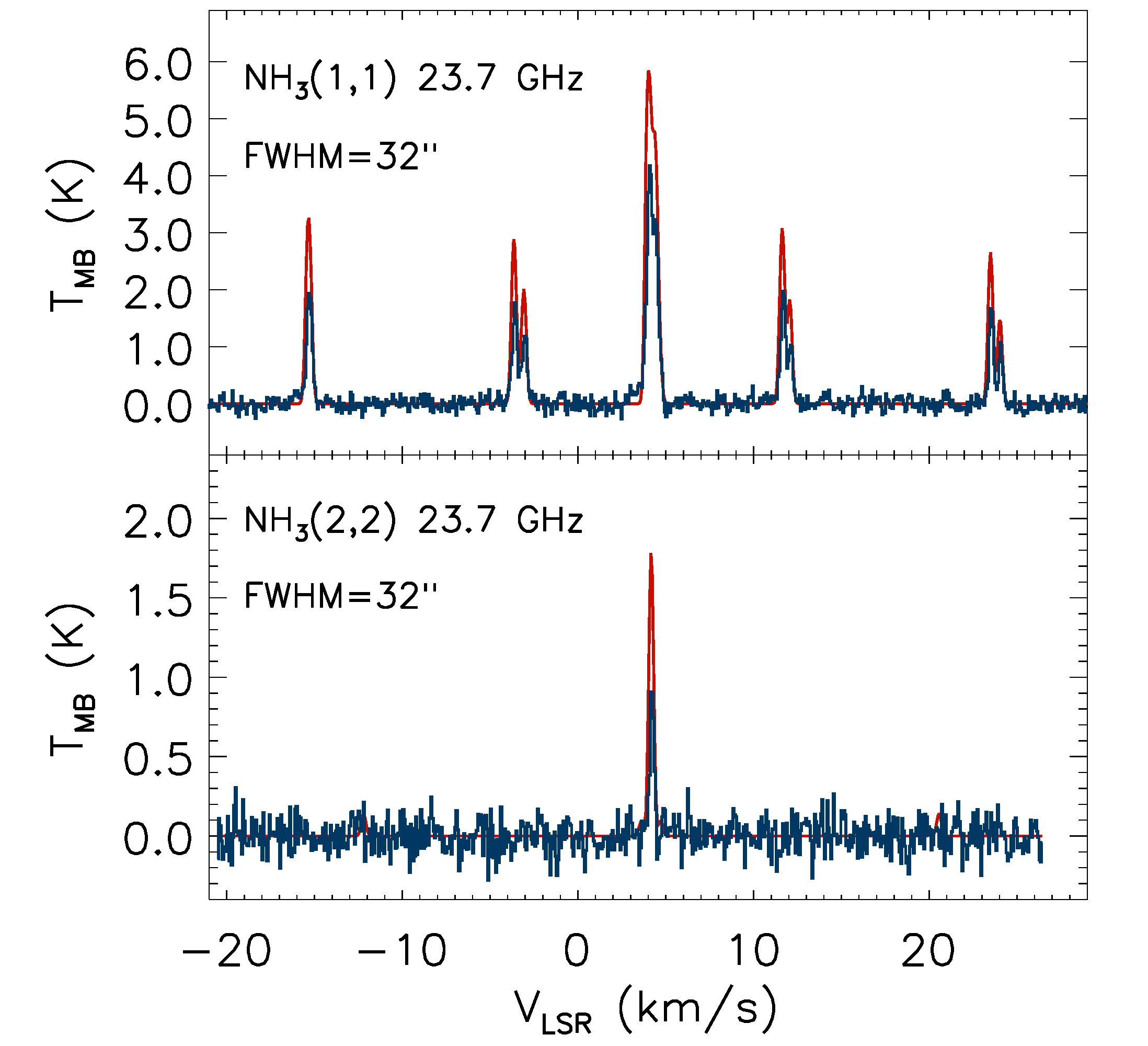}
\caption{$\ammo(1,1)$ and $(2,2)$ spectra produced by the core
  model at the time $3\,10^5$ yr (red curves) together with the
  observed spectra (histograms). The model overpredicts the observed
  intensities. The model agrees with observations at the times
  $t=10^5$ and $t=2\,10^6$ yr.  The intensities are given on the
  main-beam brightness temperature ($T_{\rm MB}$) scale.}
\label{figure:model_ammo}
\end{figure}

\begin{figure}
\includegraphics[width=9.3cm]{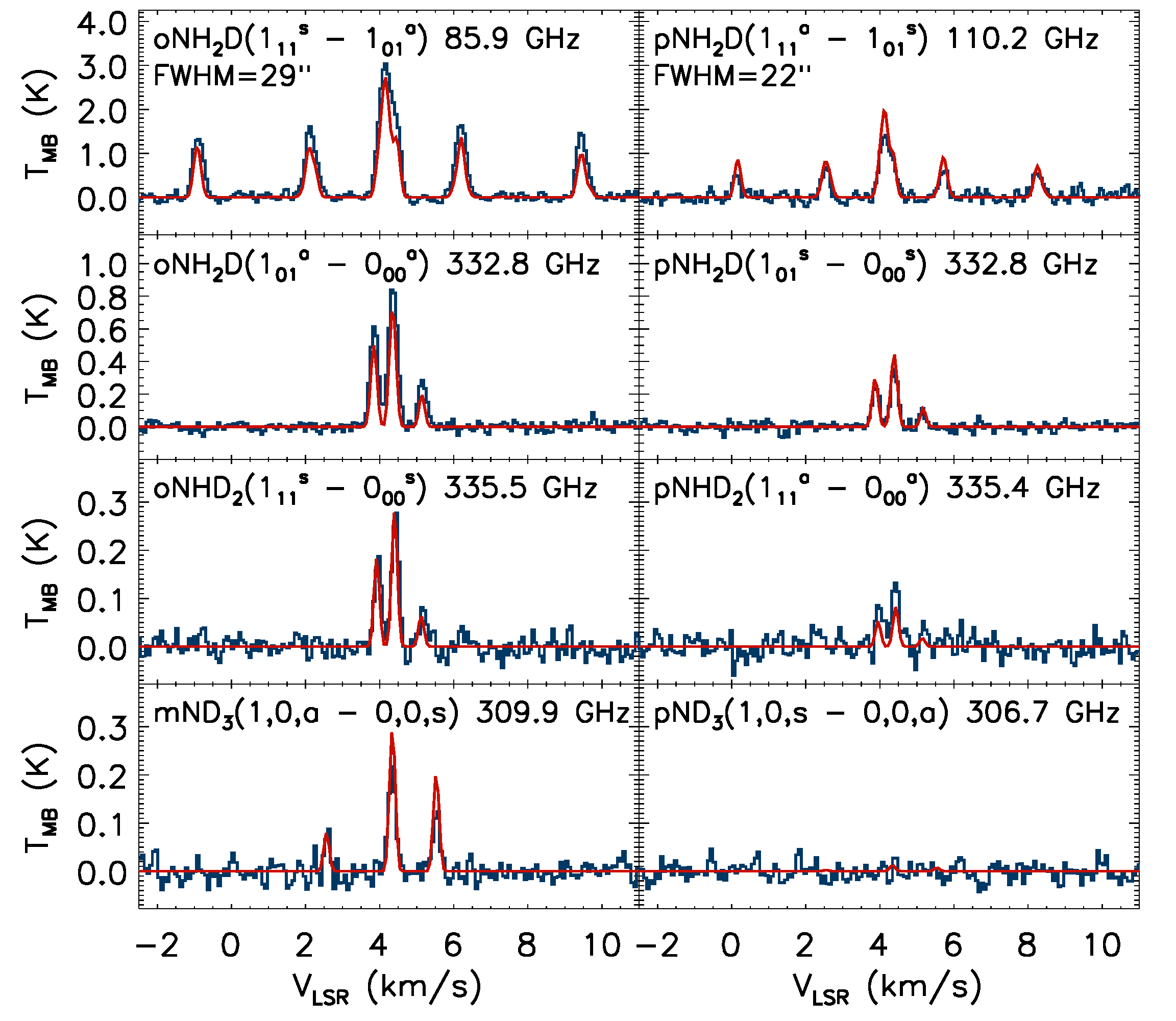}
\caption{Deuterated ammonia spectra produced by the core model at the
  time $3\,10^5$ yr (red curves) together with the observed
  spectra (histograms).}
\label{figure:model_dammo}
\end{figure}

\begin{figure}
\includegraphics[width=9.3cm]{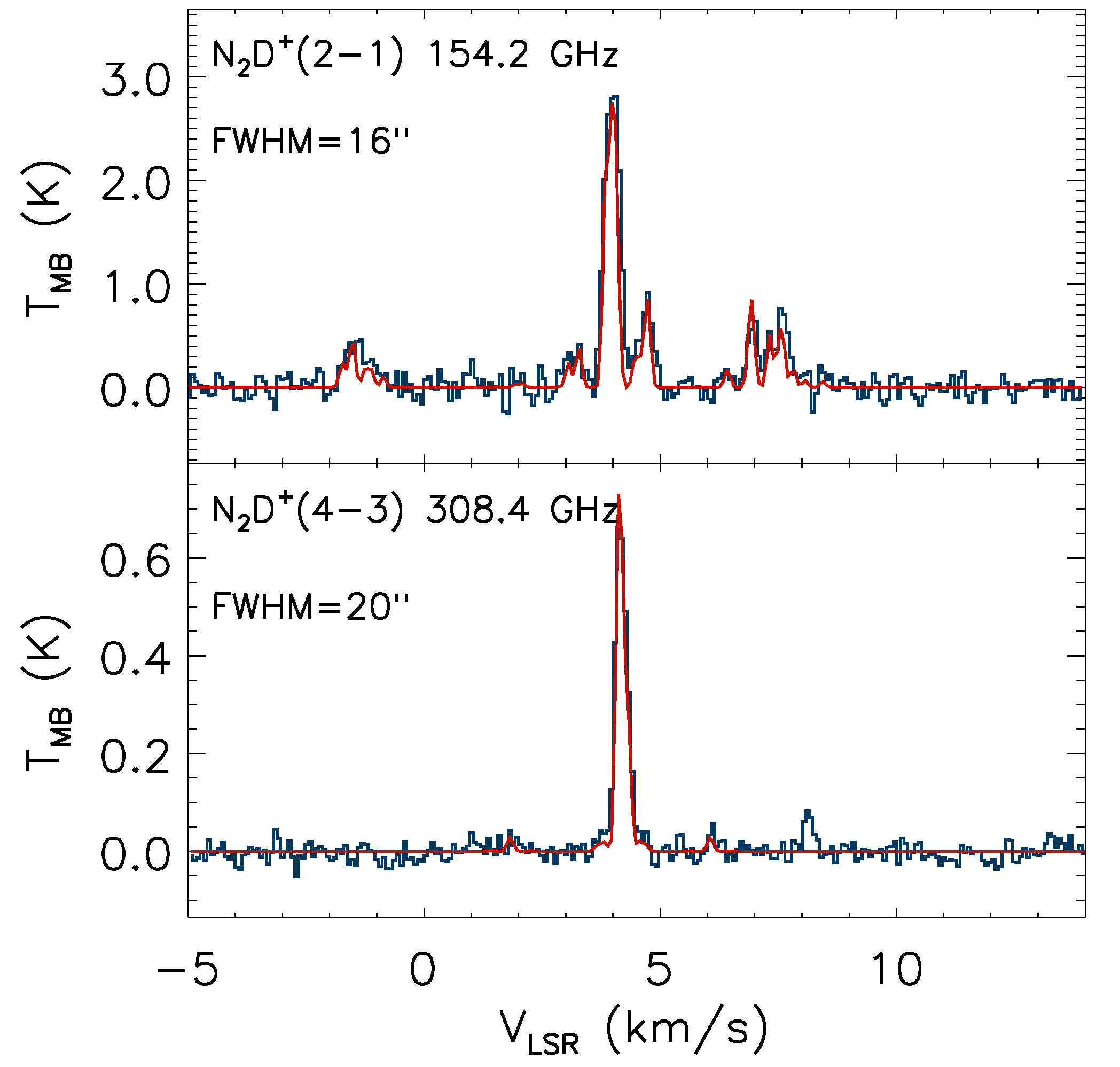}
\caption{Modelled and observed $\ddiaz(2-1)$ and $\ddiaz(4-3)$ spectra
  on the $T_{\rm MB}$ scale. The modelled spectra 
  are predictions for the time $3\,10^5$ yr (red curves) 
  after the beginning of the simulation.}
\label{figure:model_ddiaz}
\end{figure}

\begin{figure}
\includegraphics[width=9.3cm]{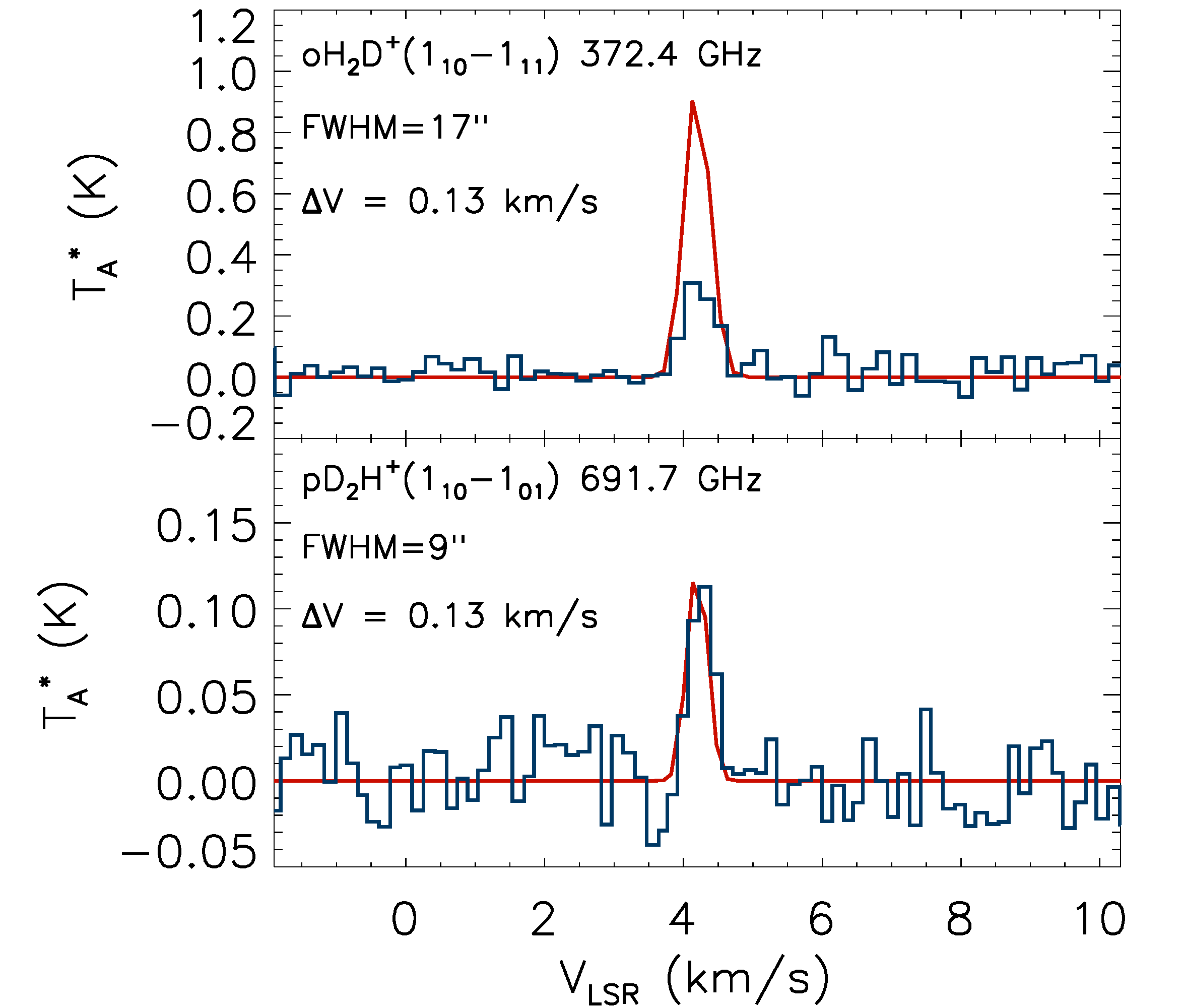}
\caption{Comparison between the observed (black histograms) and
  modelled (red curves) o$\htwod(1_{10}-1_{11})$ and
  p$\dtwoh(1_{10}-1_{01})$ spectra as observed with APEX.  The
  observations are from \cite{2011A&A...528C...2P}.  The spectra are
  on the $T_{\rm A}^*$ scale.  The model predictions are for the time
  $3\,10^5$ yr after the beginning of the simulation.}
\label{figure:oh2d+_pd2h+}
\end{figure}

The intensities of the simulated $\dammo$, $\ddammo$, $\ndthree$,
$\ddiaz$ spectra are comparable with those of the observed ones during
rather a short period around the $\ddammo/\dammo$ peak, occurring at
$\sim 3 \,10^5$ yr from beginning of the simulation. At this time,
however, the predicted beam-averaged p$\ammo$ abundance, $X({\rm
  p}\ammo)\sim5\,10^{-9}$, is about 30\% higher than needed to
reproduce the $\ammo(1,1)$ and $(2,2)$ line intensites observed at the
GBT (see Table~\ref{table:const_abus}).  
The simulated $\ammo$ lines agree with the observed ones at an
early stage, around $10^5$ yr, and much later, around $2\,10^6$ yr,
when ammonia depletion has finally started to take effect. The
predicted spectra at the time $3\,10^5$ yr are shown in
Figs.~\ref{figure:model_ammo} ($\ammo$), \ref{figure:model_dammo}
($\dammo$, $\ddammo$, and $\ndthree$), \ref{figure:model_ddiaz}
($\ddiaz$), and \ref{figure:oh2d+_pd2h+} (o$\htwod$ and p$\dtwoh$).

The model predicts that the abundances of the different isotopologues
of $\ammo$ and $\diaz$ grow at different rates.  This implies that the
fractionation ratios, $\dammo/\ammo$, $\ddammo/\dammo$,
$\ndthree/\ddammo$, do not necessarily reach their maxima at the same
time. This is illustrated in the middle panel of
Fig.~\ref{figure:abu_a_vs_time}. In this model, the peak fractionation
ratios range from 0.15 to 0.30, depending on the pair of species
considered. The $\ndthree/\ddammo$ ratio mimics the $\ddiaz/\diaz$
ratio divided by 10, and these two fractionation ratios are the first
to peak in all models we have run. This tendency is related to the
rapid growth of the $\dthree$ abundance occurring at early stages of
the simulation (Fig.~\ref{figure:abu_b_vs_time}).  The fact that the
average temperature exceeds 10 K has a favourable effect on
deuteration, but its fast advancement is made possible by the low
initial o/p$\htwo$ ratio assumed in the simulation. A higher abundance
of o$\htwo$ would delay the deuterium peak by obstructing the primary
deuteration through reaction $\hthree + {\rm HD} \leftrightarrow
\htwod + \htwo$ (\citealt{2006A&A...449..621F};
\citealt{2011ApJ...739L..35P}; \citealt{2013A&A...551A..38P};
\citealt{2015ApJ...804...98K}).

\begin{figure}
\includegraphics[width=9.3cm]{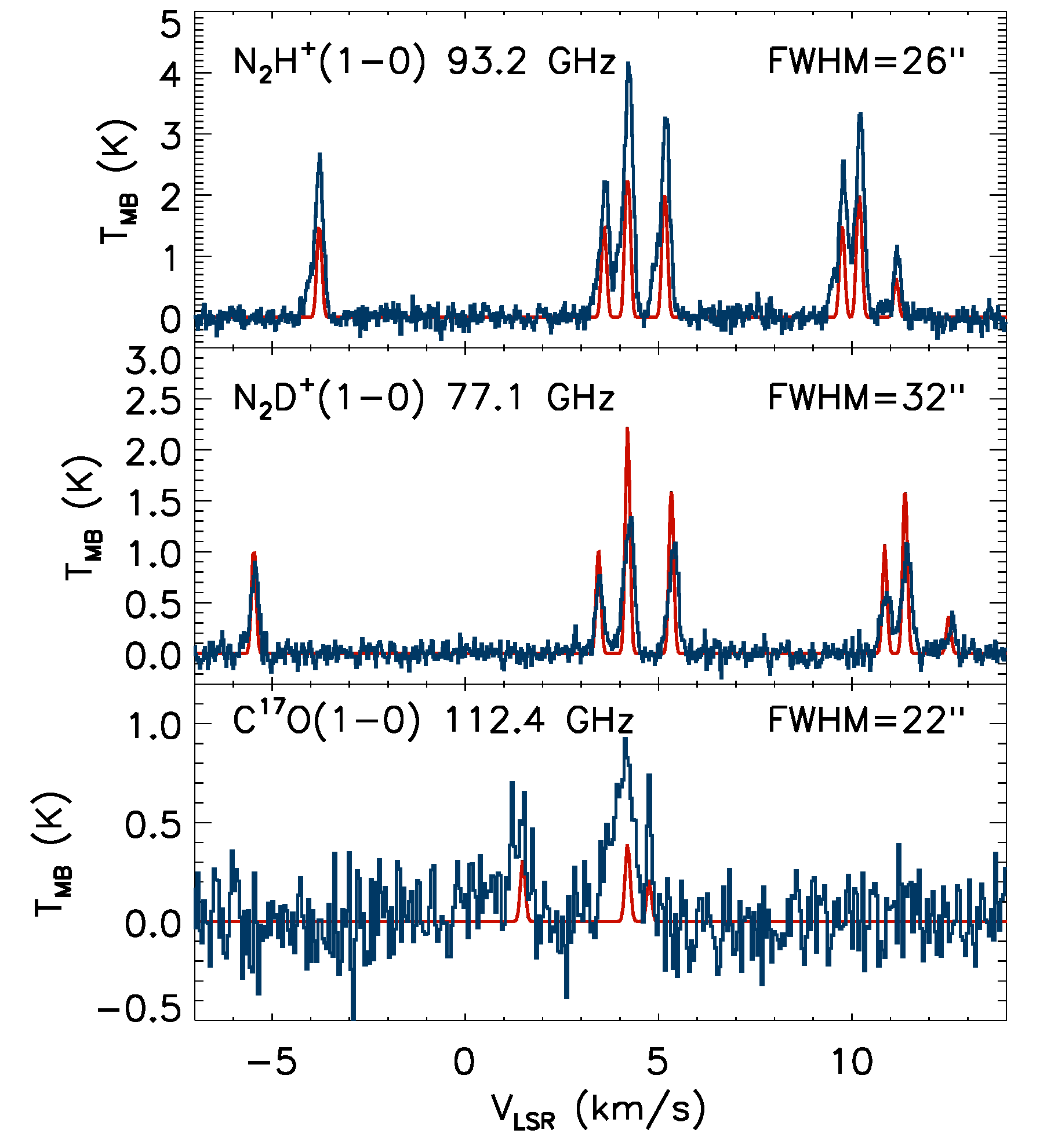}
\caption{Observed (black histograms) and modelled (red curves)
  $\diaz(1-0)$, $\ddiaz(1-0)$, and ${\rm C^{17}O}(1-0)$ spectra as
  observed with IRAM 30-m.  The observations are from
  \cite{2016A&A...587A.118P}. The spectra are on the $T_{\rm MB}$
  scale. The model spectra correspond to the time $3\,10^5$ yr in the
  simulation.}
\label{figure:n2h+_n2d+_c17o}
\end{figure}

In Figs.~\ref{figure:oh2d+_pd2h+} and \ref{figure:n2h+_n2d+_c17o} we
compare our model predictions with the previous observations of
\cite{2011A&A...528C...2P} and \cite{2016A&A...587A.118P}.  The
modelled o$\htwod(1_{10}-1_{11})$ and p$\dtwoh(1_{10}-1_{01})$ spectra
shown in Fig.~\ref{figure:oh2d+_pd2h+} are on the $T_{\rm A}^*$ scale
as observed with APEX to allow comparison with the spectra shown in
Fig.~3 of \cite{2011A&A...528C...2P}.  These spectra are reproduced in
Fig.~\ref{figure:oh2d+_pd2h+}. The $13\arcsec$ offset from the
supposed core centre position is taken into account. The hyperfine
patterns of the lines have been adopted from
\cite{1997MolPh..91..319J}. The line profiles are dominated, however,
by the thermal broadening. While the simulated p$\dtwoh$ line agrees
roughly with the observations, the simulated o$\htwod$ line is
brighter than the observed one by a factor of three. A good agreement
with both observations would be found at very early times by setting
the initial o/p$\htwo$ ratio to $10^{-4}$. On the other hand, that
model cannot reproduce the observed line ratios for the other
molecules.  After $t\sim 4\,10^5$ yr, the p$\dtwoh$ line intensity
decreases rapidly below the observed $T_{\rm A}^* \sim 0.1$ K while
o$\htwod$ remains relatively strong. This behaviour is determined by
the close correlation between o/p$\htwod$ and o/p$\htwo$, and the
increase of the o/p$\dtwoh$ ratio with time
(Fig.~\ref{figure:abu_b_vs_time}).

The $\diaz(1-0)$, $\ddiaz(1-0)$, and ${\rm C^{17}O}(1-0)$ spectra
  observed at the IRAM 30-m telescope by \cite{2016A&A...587A.118P}
  are shown in Fig.~\ref{figure:n2h+_n2d+_c17o} together with the
  predicted spectra at the time $3\,10^5$ yr. The H-MM1 spectra of
  \cite{2016A&A...587A.118P} were obtained towards the (0,0) of
  \cite{2011A&A...528C...2P}, and also here the $13\arcsec$ offset
  from the core centre is taken into account. The model reproduces the
  observed $\ddiaz(1-0)$ spectrum reasonably well, but gives 
  a vastly undervalued  ${\rm C^{17}O}(1-0)$ intensity. According to the
  model, CO is heavily depleted in the core at the time $3\,10^5$ yr
  (Fig.~\ref{figure:co_etc}). The peak velocity and velocity
  dispersion of the observed ${\rm C^{17}O}(1-0)$ line are, however,
  different from those of the $\diaz$ and $\ddiaz$ lines. The ${\rm
    C^{17}O}$ emission is probably dominated by the ambient cloud
  which is not included in our model. Also the shape of the
  $\diaz(1-0)$ line suggests that part of the emission originates in
  the ambient cloud. According to the radial abundance distributions
  shown in Fig.~\ref{figure:abu_vs_rad}, $\diaz$ belongs to species
  which are not confined to the core. Nevertherless, the model seems
  to underpredict the $\diaz(1-0)$ emission from the core.

To summarize comparisons with observations, the core model predicts
rather well the observed intensities of the $\ddiaz$, $\dammo$,
$\ddammo$, $\ndthree$, and para-$\dtwoh$ lines, overproduces
para-$\ammo$ and ortho-$\htwod$, and gives too low $\diaz(1-0)$ and 
${\rm C^{17}O}(1-0)$ line intensities. Besides these
discrepancies, the present model has problems in reproducing the
observed spin ratios. Inspection of Fig.~\ref{figure:model_dammo}
reveals that the model underestimates the o/p$\dammo$ ratio, and
overestimates the o/p$\ddammo$ ratio.

\section{Discussion}

\subsection{Ubiquity of ammonia}
\label{ss:ubiquity}

Our chemical network predicts that the ammonia abundance builds up
fast in the gas phase. The predicted fractional abundances are similar
to those found previously in molecular clouds, in particular, in the
Ophiuchus complex \citep{2009ApJ...697.1457F}, even though the model
overpredicts the p$\ammo$ abundance in H-MM1 during the deuteration
peak.  In the very beginning of the simulation, the production is
dominated by surface reactions followed by desorption, but in a few
thousand years, ion-molecule reactions in the gas phase take over as
the main source of gaseous ammonia. At early stages the rapidly
evolving carbon chemistry comes to aid as the dissociative ionization
reaction ${\rm HNC} + {\rm He^+}$ provides a short cut to NH$^+$, past
the slow formation of N$_2$ and the famous bottle-neck reaction 
${\rm N}^+ + \htwo$.
  
The ammonia abundance stays high until very late stages of the
simulation. One of the reasons is the slow decrease of the molecular
nitrogen abundance which is sustained by desorption.  In this respect
${\rm N_2}$ acts differently from CO which freezes out quickly.  In
the presence of $\hthree$ and $\hplus$, ${\rm N_2}$ replenishes the
gas with the ${\rm NH^+}$ ion through the sequence ${\rm N_2}
\arrow^{\hthree} \diaz \arrow^{\el} {\rm NH} \arrow^{\hplus} {\rm
  NH^+}$. ${\rm NH^+}$, in turn, fuels the ammonia production through
successsive reactions with $\htwo$ (see, e.g., 
\citealt{2006A&A...456..215F}; \citealt{2014A&A...562A..83L}).

In the present model, ${\rm N_2}$ attached to a grain can either
photodissociate or be desorbed. N$_2$ and CO, even if isoelectronic,
have very different surface chemistries. The two molecules have
approximately the same adsorption energies on amorphous ice: about
1100~K for CO, about 1000~K for N$_2$ (\citealt{2013ChRv..113.8783H};
\citealt{2016ApJ...816L..28F}). Hydrogenation of CO ($\mathrm{CO^*
  \rightarrow HCO^* \rightarrow H_2CO^* \rightarrow H_3CO^*\rightarrow
  H_3COH^*}$) has been experimentally observed and characterized
(\citealt{2013ChRv..113.8783H}, and references therein), at
temperatures down to 3~K, where the intermediate HCO could be detected
\citep{2011CP....380...67P}. The reaction ${\rm CO} + {\rm H}
\rightarrow {\rm HCO}$ proceeds by tunneling through an activation
barrier computed to be about 2000~K in the gas phase, and probably
lower on the surface of amorphous ice (\citealt{2013JChPh.139p4310P};
\citealt{2014A&A...572A..70R}).  The exothermicity of the reaction is
about 6700~K, while the endothermicity of the reaction ${\rm CO} +
{\rm H} \rightarrow {\rm COH}$ is about 10000~K
(\citealt{2007JChPh.126r4308Z}; note that this theoretical paper
overestimates the exothermicity of the reaction leading to HCO). The
case of N$_2$ is different: The reaction $\mathrm{N_2 + H \rightarrow
  N_2H}$ is endothermic by about 4400~K ~\citep{2010JChPh.132f4308B}
and does not occur on ice surfaces. The CO and N$_2$ neutral chemistry
proceed thus in very different ways, with no surface hydrogenation of
N$_2^*$ towards ammonia NH$_3$ or hydrazine
$\mathrm{N_2H_4}$. However, the ionic chemistry on low temperature ice
surfaces is not fully characterized.
    
\subsection{Fractionation ratios}
\label{ss:fractionation}

\begin{figure}[hbt]
\centering
   \unitlength=1.0mm
  \begin{picture}(90,22)(0,0)
\color{blue} 
\put(3,13){\makebox(0,0){$\ammo$}}

\color{black} 

\put(7,13){\vector(1,0){18}}
\put(16,15){\makebox(0,0){$\htwod$,$\dtwoh$}}

\put(31,13){\makebox(0,0){$\dammohplus$}}

\put(35,13){\vector(1,0){19}}
\put(45,15){\makebox(0,0){$\el$}}

\color{blue} 
\put(59,13){\makebox(0,0){$\dammo$}}

\color{black} 
\put(64,13){\vector(1,0){17}}
\put(73,15){\makebox(0,0){$\hplus$}}

\put(59,10){\vector(0,-1){5}}
\put(59,3){\makebox(0,0){grains}}

\put(86,13){\makebox(0,0){$\dammoplus$}}
\put(86,15){\line(0,1){5}}
\put(86,20){\line(-1,0){56}}
\put(30,20){\vector(0,-1){5}}
\put(59,22){\makebox(0,0){$\htwo$}}
\color{black}
\end{picture}
\caption{Principal reactions forming and destroying $\dammo$ at 
the deuteration peak.}
      \label{figure:dammo_formation}
\end{figure}
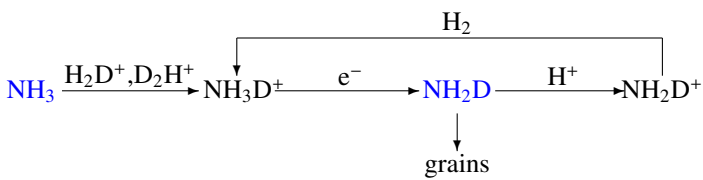

The abundance of $\dammo$ starts to increase gradually, first through
the deuteron transfer to ammonia, primarily by ${\rm HCND^+}$ or ${\rm
  DCNH^+}$, e.g., ${\rm HCND^+} + \ammo \rightarrow \dammohplus + {\rm
  HCN}$, followed by dissociative recombination of $\dammohplus$.  The
depletion of CO boosts the abundance of $\hthree$, which in turn is
efficiently deuterated to $\htwod$, $\dtwoh$, and $\dthree$ in
successive reactions with HD. This stage is characterised by a rapid
increase of $\ddammo$, $\ndthree$, and $\ddiaz$.  The most important
reactions contributing to the formation of $\dammo$ during the
deuteration peak are shown in Fig.~\ref{figure:dammo_formation}. These
comprise the deuteron transfer from $\htwod$, or some other deuterated
ion, to $\ammo$ giving $\dammohplus$, dissociative recombination of
$\dammohplus$, charge transfer between $\dammo$ and ${\rm H^+}$, and
hydrogen abstraction from $\htwo$ to the $\dammoplus$ ion.
Characteristic of the reaction scheme is the circulation between
neutral and ionic species generated by the charge transfer reaction
with $\hplus$, one of the ions which increase after the disappearance
of CO. Precursors of doubly and triply deuterated ammonia,
$\ddammohplus$ and $\ndthreehplus$, are mainly formed from reactions
between $\ammo$ and $\dtwoh$ or $\dthree$ in the present model.  This
reaction network is discussed in more detail in
\cite{2015A&A...581A.122S}.

The $\ddiaz$ ion, which is mainly produced in reactions between ${\rm N_2}$
and $\htwod$, $\dtwoh$, or $\dthree$, is strongly favoured by the
successive deuteration of $\hthree$. At the time of the most vigorous
deuteration, the $\ammo$ abundance decreases slightly owing to
accretion onto grains and enhanced charge exchange reactions caused by
the increase of ${\rm H^+}$. After that, until very late times, 
probably exceeding the lifetime of the core, the $\ammo$ abundance
remains almost constant, and so does the abundance of $\dammo$. In
contrast, the abundances of $\ddammo$, $\ndthree$, $\diaz$, and $\ddiaz$
rise and fall in the time range shown in
Fig.~\ref{figure:abu_a_vs_time}. At late times these species are most
strongly influenced by the depletion of nitrogen and deuterium in the
gas phase.

The best overall agreement between the modelled and observed 
 $\dammo$, $\ddammo$, and $\ndthree$ spectra is achieved at the time
$\sim 3\,10^5$ yr after beginning of the simulation. At this
  stage of the model, the fractionation ratios are $\dammo/\ammo \sim
  \ddammo/\dammo \sim 0.25$, and $\ndthree/\ddammo \sim 0.1$ (while
  the observed fractionation ratios are $\dammo/\ammo \sim 0.4$,
  $\ddammo/\dammo \sim 0.2$, and $\ndthree/\ddammo \sim 0.06$).
The time is coincident with the $\ddammo$ maximum, whereas $\ndthree$ is
already going down then.  The obtained fractionation ratios are
reasonably close to what is found in previous observational studies
(\citealt{2005A&A...438..585R}; \citealt{2016MNRAS.457.1535D}), but it
should be noted that our results suggest that large temporal
variations are possible, also when the physical conditions remain
constant. The spectral line simulations show that all three deuterated
forms of ammonia should be easily detectable from a core like H-MM1
even if the fractionation ratios where reduced to half of those
derived here. In particular, the early formation of $\ndthree$ and the
large transition dipole moment of the rotation-inversion transition of
m$\ndthree$ at 309.9 GHz makes this line a useful signpost of the
deuterium peak.

According to the present chemistry model, the fractionation ratios on
grain surfaces are lower than those in the gas phase by a factor of
two. The atomic D$^*$/H$^*$ ratio on the grain surfaces, which
determines the overall degree of deuteration there, reaches a high value of
$\sim 0.4$ a little before $10^5$ yr in the present simulation
(Fig.~\ref{figure:abu_c_vs_time}).  Because of competition between
various viable addition reactions for H$^*$ and D$^*$ (for example
with NO$^*$, HCO$^*$, and HS$^*$), the abundances of deuterated forms
of ammonia build up slowly. In the end of the simulation 
the abundances settle, however, almost exactly to the values expected
from the statistical
rule $\dammo^*/\ammo^* = \frac{3}{\sqrt{2}} {\rm D^*/H^*}$, 
$\ddammo^*/\dammo^* = \frac{1}{\sqrt{2}} {\rm D^*/H^*}$, 
$\ndthree^*/\ddammo^* = \frac{1}{3\sqrt{2}}
{\rm D^*/H^*}$, as predicted by \cite{1989MNRAS.240P..25B}.

\subsection{Spin ratios}
\label{ss:spinratios}
 
In the gas phase, the {\sl ortho/para} ratio of $\dammo$ is largely
determined by the cycle consisting of reactions with $\hplus$,
$\htwo$, and $\el$ shown in Fig.~\ref{figure:dammo_formation}.  In the
present chemistry model, full scrambling of H nuclei is assumed to
take place in these reactions, and, owing to nuclear spin selection
rules, o/p$\dammo$ should settle to about 2.3
\citep{2015A&A...581A.122S}. The full reaction set predicts that the
ratio decreases to about 2.0 at late times
(Fig.~\ref{figure:abu_a_vs_time}).

Similar cycles involving $\ddammo$ and $\ndthree$ preserve the spin
states of ${\rm D_2}$ and ${\rm D_3}$. Consequently, the spin ratios
of doubly and triply deuterated ammonia are determined by the primary
deuteration reactions $\dtwoh + \ammo$ and $\dthree + \ammo$.  The
reaction scheme is discussed in detail in
\cite{2015A&A...581A.122S}. The fact that o/p$\ddammo$ follows closely
o/p$\dtwoh$ can be seen in Figs.~\ref{figure:abu_a_vs_time} and
\ref{figure:abu_b_vs_time}. A tight correlation between m/p-$\ndthree$
and m/p-$\dthree$ is also evident from these figures.

By comparing the modelled and observed spectra shown in
Fig.~\ref{figure:model_dammo} one finds that while the modelled
o/p$\dammo$ ratio is too low, the corresponding ratio for $\ddammo$ is
too high. The discrepancy is more pronounced in the case of p$\ddammo$
for which the predicted spectrum underestimates the observed intensity
by about 40\%. As mentioned in Sect.~\ref{ss:chem_evolution}, the
timing and the strength of the deuterium peak can be affected by the
selection of the initial o/p-$\htwo$ ratio, the cosmic rays ionization
rate, and the average grain size, but these modifications do not
change the fact that the o/p$\ddammo$ ratio given by our model during
the deuterium peak is larger than the observed ratio. If the adopted
deuteration scheme is correct, the implication is that also the
o/p$\dtwoh$ ratio is overestimated in the model, because o/p$\ddammo$
is directly related to o/p$\dtwoh$.

However, the spin ratio of $\dtwoh$ is determined by the well-studied
$\hthree + \htwo$ isotopic system \citep{2009JChPh.130p4302H}, and the
prediction of a high o/p$\dtwoh$ ratio seems to be well-founded. The
lower energy, {\sl ortho} form of $\dtwoh$ is favoured over p$\dtwoh$
both in the primary production through $\htwod + {\rm HD} \rightarrow
\dtwoh +\htwo$ and in the ``backward'' reactions ${\rm m/o}\dthree +
{\rm o}\htwo \rightarrow {\rm o/p}\dtwoh + {\rm HD}$ and ${\rm
  o/p}\dtwoh + {\rm o}\htwo \rightarrow {\rm o/p}\htwod + {\rm
  HD}$. In addition, {\sl para - ortho} conversion of $\dtwoh$ is
viable in cold clouds through reaction with HD (see discussion in
\citealt{2006A&A...449..621F} and \citealt{2010A&A...509A..98S}).

\subsection{Statistical abundance ratios}
\label{ss:statistical}

In their analysis of the deuterated ammonia observations towards
Barnard 1 and L1689N, \cite{2016MNRAS.457.1535D} concluded that the
observational data (considering the error margins) are consistent with
the assumption that both the fractionation ratios and the spin ratios
are equal to the corresponding statistical ratios. 
\cite{2016MNRAS.457.1535D} point out that statistical
fractionation and spin ratios would be expected if the production of
ammonia molecules were dominated by surface reactions. On grain surfaces,
where ammonia formation takes place through H/D atom additions to N,
the fractionation ratios, $\dammo/\ammo$, $\ddammo/\dammo$, and
$\ndthree/\ddammo$, should successively diminish by a factor of three
(\citealt{1989MNRAS.240P..25B}; \citealt{2001ApJ...553..613R}), and
the spin ratios should follow the ratios of the corresponding nuclear
spin statistical weights.

Also in the present study, the nuclear spin ratios derived
  directly from the observed lines agree with their statistical
  values, and $\ddammo/\dammo \sim 3 \ndthree/\ddammo$ as
  expected from combinatorial principles (see
  Table~\ref{table:const_abus} in Sect.~\ref{ss:ave_abus}). The
  $\dammo/\ammo$ ratio falls, however, below the value expected from
  the other two fractionation ratios. On the other hand, the derived
  $\ammo$ abundance for the core may be an overestimate as discussed
  in Sect.~\ref{ss:ave_abus}.

The observations pertain, however, gas-phase molecules, and also
sublimated species are likely to be exposed to rapid processing by
ion-molecule reactions in the gas phase. The fractionation ratios on
grain surfaces depend on the atomic D$^*$/H$^*$ ratio on grains, which
according to the present simulation is at any time much higher that
the atomic D/H ratio in the gas phase (Figs.~\ref{figure:co_etc} and
\ref{figure:abu_c_vs_time}). As mentioned in
Sect.~\ref{ss:fractionation}, the statistical fractionation ratios for
ammonia on grains are only reached at very late times of the
simulation.

The observations suggest therefore that the adopted gas-phase
deuteration scheme for ammonia (illustrated in Fig.~2 of
\citealt{2015A&A...581A.122S}) is not correct.  The model assumes
complete scrambling of H or D nuclei in the intermediate reaction
complexes. This assumption is highly uncertain for several reactions
involved in the production of ammonia \citep{2013JPCA..117.9800R}.

An argument against a full scrambling of the reaction forming the
ammonium ion, $\ammoplus + \htwo \rightarrow (\mathrm{NH_5^+})^\ddag
\rightarrow \ammohplus + {\rm H}$, comes from the energetics of this
reaction (an intermediate reaction complex is indicated here with
$()^\ddag$). The reaction is likely to occur through two minima on
the potential energy surfaces, $(\ammo\cdot\cdot\htwo^+)^\ddag$ and
$(\mathrm{NH_4}\cdot\cdot\mathrm{H}^+)^\ddag$. According to the
calculations of \cite{1992JChPh..97.1191I}, some of the conceivable
interchanges of two H nuclei between different parts of these
complexes are facile, but most of them involve high energy
$(\mathrm{NH_5^+})^\ddag$ transition states with different
geometries. The D-substituted cases should be very similar.  However,
since it is known experimentally that, for example, D/H exchange does
occur between $\ammoplus$ and ${\rm D_2}$, \citet{1992JChPh..97.1191I}
proposed that it is possible to circumvent a high-energy transition
state by a process where hydrogen transfer is followed by internal
rotation and reverse transfer.  In none of the configurations of
$(\mathrm{NH_5^+})^\ddag$ do all H (or D) nuclei occupy equivalent
positions. Consequently, the spin symmetry rules for this reaction are
far from obvious, all the more that internal rotations are likely to
occur.

For the reaction $\mathrm{NH_3 + H_2D^+}$ and its doubly deuterated
analogue (Fig.~\ref{figure:ammo+dtwoh} below), the situation is even
less clear. While the reaction proceeds at 30~K
(\citealt{1975JChPh..62.3549L}; \citealt{1989A&A...213L..29M}), there
is no theoretical computation on the $(\mathrm{NH_5D^+})^\ddag$
complex, see for example \citet{2013JPCA..117.9800R}. There is at the
moment no experimental or theoretical discrimination between
scrambling or proton hop mechanisms for this reaction.

In the present model we have assumed that the reaction $\ammo +\dtwoh$
does form the complex $({\rm NH_4D_2^+})^\ddag$ which can dissociate to
$\ammohplus$, $\dammohplus$, or $\ddammohplus$. If this is true, the
reaction is one of the main sources of $\ddammo$.  If the reaction
complex is not formed, but $\dtwoh$ just donates the proton or one of
the deuterons to ammonia, the outcome after dissociative recombination
can be either $\ammo$ or $\dammo$, as illustrated in
Fig.~\ref{figure:ammo+dtwoh}. The statistical branching ratios are
indicated in this figure.

\begin{figure}[hbt]
\centering
\unitlength=1.0mm
\begin{picture}(80,27)(0,0)
\color{black} 

\put(7,15){\makebox(0,0){$\ammo+\dtwoh$}}

\put(45,25){\makebox(0,0){${\rm NH_4^+}$}}

\put(45,5){\makebox(0,0){${\rm NH_3D^+}$}}

\put(80,25){\makebox(0,0){$\ammo$}}

\put(80,5){\makebox(0,0){$\dammo$}}

\put(18,15){\vector(2,1){20}}
\put(18,15){\vector(2,-1){20}}
\put(25,23){\makebox(0,0){1/3}}
\put(25,7){\makebox(0,0){2/3}}

\put(50,25){\vector(1,0){25}}
\put(50,5){\vector(1,0){25}}
\put(50,7){\vector(3,2){25}}

\put(60,27){\makebox(0,0){1/1}}
\put(60,17){\makebox(0,0){1/4}}
\put(60,2){\makebox(0,0){3/4}}
\end{picture}

\caption{Branching ratios of the reaction $\ammo + \dtwoh$ 
  assuming that this can be described as proton/deuteron hop.}
\label{figure:ammo+dtwoh}
\end{figure}
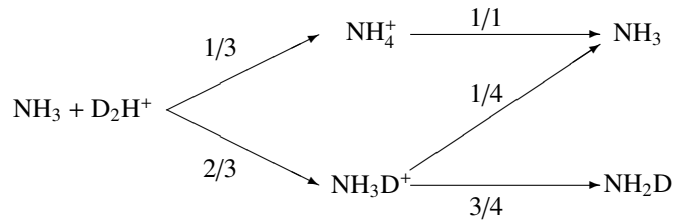

It seems that several important reactions which in the model have been
assumed to proceed through long-lived intermediate complexes where
nuclei can be scrambled, should rather be described as proton/deuteron
hops or hydrogen/deuterium abstractions. In these reactions H and D
nuclei are effectively added one by one, like in surface reactions,
and they produce each nuclear spin modification according to its
statistical weight.

The assumption that ammonia is primarily processed in reactions with
the isotopologues of $\hthree$, like the one described in
Fig.~\ref{figure:ammo+dtwoh}, can eventually lead to a ratio of 3 between
successive levels of deuteration, although now the fractionation
ratios would not depend on D/H nor D$^*$/H$^*$, but on the relative
abundances of $\hthree$, $\htwod$, $\dtwoh$, and $\dthree$. Using
combinatorics one can show that in steady state $\dammo/\ammo = 3\gamma$,
$\ddammo/\dammo = \gamma$, and $\ndthree/\ddammo =\frac{1}{3}\gamma$,
where
$$
\gamma=\frac{[\htwod]+2[\dtwoh]+3[\dthree]}{3[\hthree]+2[\htwod]+[\dtwoh]}\;.
$$ 
 
\cite{2016MolAs...3...10N} has recently discussed a simple statistical
model for the deuteration of interstellar ammonia which neglects the
reaction kinetics. In its simplest form this model leads to the same
rule for the fractionation ratios, $\dammo/\ammo = 3\,\ddammo/\dammo =
9\, \ndthree/\ddammo$, as the models discussed above, but now the
parameter $\gamma = \ddammo/\dammo$ depends on the elemental N/D ratio
through $\gamma = 1/(3\,{\rm N/D} - 1)$.  The fractionation
ratio $\ddammo/\dammo \sim 0.2$ observed in H-MM1 and in L1689N
\citep{2005A&A...438..585R} would imply N/D$\sim 2$, which is about
30\% lower than the ratio assumed in the chemistry model used
here. The inclusion of energetics into this model through the
molecular partition functions makes the fractionation ratios to
correspond to what would be obtained at local thermodynamic
equilibrium, giving $\dammo/\ammo \leq \ddammo/\dammo$. While this
agrees with the earlier results presented in
\cite{2005A&A...438..585R}, the fractionation ratios found in the
present study, and in the re-analysis of the B1b results by
\cite{2016MNRAS.457.1535D}, suggest $\dammo/\ammo \sim 2-3\,
\ddammo/\dammo$. Therefore, the assumption of thermal equilibrium
between different deuterated isotopologues of ammonia does not seem to
be universally valid. This situation is also unlikely because large
deviations from thermal equilibrium are usually found in cold
interstellar gas. On the other hand, as recommended by
\cite{2016MolAs...3...10N}, the energetics, and especially the
differences in the vibrational zero-point energies between different
ammonia isotopologues should be taken into account in kinetic models.
This would affect in particular reactions working against deuteration.

\section{Conclusions}

Ammonia and its three deuterated isotopologues were detected towards
the starless core H-MM1 in Ophiuchus. By modelling the observed
spectra we derived the following fractionation ratios:
$\dammo/\ammo\sim0.4$, $\ddammo/\dammo \sim 0.2$,
$\ndthree/\ddammo\sim 0.06$.  The relative line intensities of the
{\sl ortho} and {\sl para} modifications of $\dammo$ and $\ddammo$ are
consistent with the statistical spin ratios o/p$\dammo=3$,
o/p$\ddammo=2$. The fractionation and spin ratios are similar to
those obtained towards two young cores (L1689N and B1), which are the
only objects observed previously in all four molecules
(\citealt{2005A&A...438..585R}; \citealt{2015A&A...576A..99R};
\citealt{2016MNRAS.457.1535D}).

The observations towards H-MM1 were simulated using a gas-grain
chemistry model in conjunction with a Monte Carlo radiative transfer
program. In the chemistry model, which includes cosmic-ray induced
desorption and tunnelling through activation energy barriers in
surface reactions, ammonia forms early and stays long in the gas
phase.  The model overpredicts p$\ammo$ and o$\htwod$, but can
approximately reproduce the observed $\dammo$, $\ddammo$, $\ndthree$,
$\ddiaz$, and p$\dtwoh$ lines. The o$\htwod$ and p$\dtwoh$
observations used here are from \cite{2011A&A...528C...2P}. According
to the simulation, the longevity of $\ammo$ and $\diaz$ in the gas
phase can be traced back to the chemical inertness of ${\rm N_2}$ on
the grain surfaces.  Unlike CO, the nitrogen molecule is not supposed
to be hydrogenated on grain surfaces, which makes it susceptiple 
to desorption. 

The present chemistry model cannot account for the observed
o/p$\dammo$ and o/p$\ddammo$ ratios satisfactorily. In conditions
prevailing in H-MM1 and in other starless, dense cores, characterised
by a low temperature, high obscuration, and inefficient desorption,
the spin ratios should be determined by gas-phase ion-molecule
reactions. When complete scrambling of H and D nuclei in these
reactions is assumed, the spin ratios are predicted to settle to
o/p$\dammo\sim 2$ and o/p$\ddammo\sim 3-4$ by the time all three
deuterated forms of ammonia become detectable, instead of their
statistical ratios 3 and 2, respectively. The fact that the observed
spin ratios nevertheless correspond to the nuclear spin statistical
weights suggests that full scrambling in reactions forming
deuterated ammonia is not a valid assumption. At the moment there are
very little experimental data and few theoretical calculations
concerning the probability of proton/deuteron scrambling in the
principal deuteration reactions of ammonia.

In constrast to what our chemistry model predicts, the currently
available observational data suggest that the nuclear spin ratios of
deuterated ammonia isotopologues do not depend strongly on physical
conditions or time. On the other hand, the degree of deuterium
fractionation does show clear variations from source to source, and is
also predicted to have a temporal pattern, depending on CO depletion,
the ortho/para-$\htwo$ ratio, and eventually on HD depletion. The
deuterium fractionation of ammonia has therefore the potential to be
used as a probe of dense core evolution. While the present core model
is static, chemical reaction rates depend strongly on the density,
i.e., the dynamical evolution. Therefore definitive conclusions on the
effects of cloud evolutionary stages on deuterium fractionation awaits
for models where deuterium chemistry is coupled to the dynamical
evolution.

\begin{acknowledgements}
  
  We thank the anonymous referee for insightful comments which helped to
  improve the manuscript. We thank Malcolm Walmsley and Juris
  Kalv{\=a}ns for helpful discussions.  J.H. and L.W. thank the
  Max-Planck-Institute for Extraterrestrial Physics for generous
  support. J.H. acknowledges financial support from the Academy of
  Finland grant 258769. P.C., J.P., and A.P. acknowledge the financial
  support of the European Research Council (ERC; project PALs
  320620). F.D., L.W., A.F., and C.R. thank the Agence Nationale de la
  Recherche (ANR-HYDRIDES), contract No. ANR-12-BS05-0011-01, and the
  CNRS national program "Physique et Chimie du Milieu Interstellaire".

\end{acknowledgements}

\bibliographystyle{aa}
\bibliography{harju_aa_2016_28463}

\end{document}